\documentclass[a4paper,fleqn,usenatbib]{mnras}

\usepackage[T1]{fontenc}
\usepackage{ae,aecompl}
\usepackage{newtxtext,newtxmath}

\usepackage{graphicx}	
\usepackage{amsmath}	
\usepackage{pdflscape}
\newcommand{\ha}{H$\alpha$~}
\newcommand{\OIII}{[O{\sc{iii}}]}
\newcommand{\OII}{[O{\sc{ii}}]}
\newcommand{\NII}{[N{\sc{ii}}]}
\newcommand{\SII}{[S{\sc{ii}}]}
\newcommand{\HII}{H{\sc{ii}}}

\title[Metallicity conversions in MaNGA]{Conversions between gas-phase metallicities in MaNGA}

\author[J.M. Scudder et al.]{
Jillian M. Scudder,$^{1}$\thanks{E-mail: jillian.scudder@oberlin.edu}
Sara L. Ellison,$^{2}$
Loubna El Meddah El Idrissi$^{1}$, and Henry Poetrodjojo$^{3, 4}$\\
$^{1}$Department of Physics \& Astronomy, Oberlin College, Oberlin, OH, 44074, USA \\
$^{2}$Department of Physics \& Astronomy, University of Victoria, Finnerty Road, Victoria, British Columbia, V8P 1A1, Canada\\
$^{3}$Research School of Astronomy and Astrophysics, The Australian National University, Cotter Road, Weston, ACT 2611, Australia\\
$^{4}$ARC Centre of Excellence for All Sky Astrophysics in 3 Dimensions (ASTRO 3D)\\
}

\date{Accepted XXX. Received YYY; in original form ZZZ}

\pubyear{2021}

\begin{document}
\label{firstpage}
\pagerange{\pageref{firstpage}--\pageref{lastpage}}
\maketitle

\begin{abstract}
We present polynomial conversions between each of 11 different strong line gas-phase metallicity calibrations, each based on $\sim$ 1.1 million star-forming spaxels in the public Sloan Digital Sky Survey (SDSS) Data Release 15 (DR15) Mapping Nearby Galaxies at Apache Point Observatory (MaNGA) survey. For this sample, which is $\sim$ 20 times larger than previous works, we present 5th order polynomial fits for each of 110 possible calibration conversions, for both Small Magellanic Cloud (SMC)-type and Milky Way (MW)-type dust corrections. The typical $2\sigma$ scatter around our polynomial fits is 0.1 dex; we present the range over which the metallicities are valid.
Conversions between metallicities which rely on the same set of line ratios, or a heavily shared set of emission lines, have reduced scatter in their conversions relative to those conversions with little overlap in required emission lines. Calibration conversions with less consistent sets of emission lines also have increased galaxy-to-galaxy variability, and this variability can account for up to 35\% of the total scatter. 
 We also compare our conversions to previous work with the single fibre SDSS DR7 spectra along with higher spatial resolution data from the TYPHOON Integral Field Spectroscopy survey, resulting in comparison samples with spatial resolutions from several kpc down to $\sim$100 pc. Our metallicity conversions, obtained with the large sample of MaNGA, are robust against the influence of diffuse ionized gas, redshift, effective radius and spatial blurring, and are therefore consistent across both integrated spectra and the high resolution integral field spectroscopy data.  
\end{abstract}




\section{Introduction}

In looking to find useful metrics by which we can trace a galaxy's evolution through cosmic time, gas-phase metallicity has been used for many years as a tracer of the enrichment of the interstellar medium by generations of stellar end-of life cycles. Whether by the dramatic, highly energetic enrichment events of a supernova of a high mass star, or the longer timescale process of a solar mass star sloughing off its outer layers into a planetary nebula, it is the stars of a galaxy which build hydrogen up into heavier elements.  To find low metallicity gas in the local Universe is therefore to find gas which has not cycled through the core of very many stars; conversely, high metallicity gas has been enriched by the journey through many generations of stars. 

Through the measurement of gas-phase metallicities via emission lines, it has been found that galaxies typically have higher metallicities in their centres and lower metallicities at larger galactic radii \citep[e.g.,][]{Rich2012, Sanchez-Menguiano2016}, that there exists a strong relationship between the stellar mass of a galaxy and its metallicity \cite[e.g.,][]{Lequeux1979, Tremonti2004, Lee2006, Kewley2008, Lara-Lopez2010, Sanchez2014, Sanchez2019}, and that these relationships persist with a change in scaling out to high redshift \citep[e.g.,][]{Kobulnicky2003, Liang2006, Maiolino2008, Mannucci2010, Zahid2011, Maier2014a, Sanders2020}. 

Deviations from these trends found in the general galaxy population can be used to infer the existence of some perturbing event, such as a recent inflow of low metallicity gas.  
At low redshift, flattened metallicity gradients could serve as a flag of a recent or ongoing merger event which has perturbed the gravitational potential of the galaxy, as suggested by both theoretical \citep[e.g.,][]{Mihos1996, Rupke2010a, Torrey2012, Bustamente2018} and observational results \citep[e.g.,][]{Ellison2008, Rupke2010b, Kewley2010a, Scudder2012b, Sanchez2014a, Barrera-Ballesteros2015}, or as evidence for a galactic scale fountain of gas redepositing high metallicity gas in the outskirts of the galaxy \citep[e.g.,][]{Belfiore2014, Belfiore2017}. 
Atypically high central metallicities may also serve to trace galactic environments where such perturbing events are rare \citep[e.g.,][]{Scudder2012}. 

Accurate measurements of gas-phase metallicities are therefore very helpful for studies of galaxy evolution, but these measurements are not always straightforward. Fundamentally we are attempting to assess the metal content within a cloud of gas which is dominated (if we count by individual nuclei) by hydrogen, but any metric we develop to attempt this must account for (or be insensitive to) physical changes in gas temperature, ionization parameters, and density, each of which can change the strength of the emission lines from the metals we wish to trace \citep[e.g.,][]{Kewley2019}. 

There is a direct method of measuring the gas-phase metallicity, which is to use two different ionization species of the same element. With a ratio of lines with two different ionization levels (usually \OIII$\lambda4363$ and the lower energy \OIII$\lambda5007$), the gas temperature can be directly calculated, and a metallicity inferred by solving for this gas temperature \citep{Liang2007, Andrews2013}. However, the higher energy \OIII$\lambda4363$ forbidden line required for this direct method is typically very weak in all except for very specific physical conditions - high gas temperature, low density, and low metallicity \citep{Kewley2019}. The limitations on the direct method mean that to use it as a probe of higher metallicity, lower gas temperature, or higher density regions, spectral stacking is typically  required \citep{Liang2007, Andrews2013,Brown2016, Bian2018, Curti2020}. 

These limitations have driven the development of other methodological pathways to calculate gas-phase metallicity. The most common alternative is to use calibrations of strong emission line ratios. These strong emission lines are much easier to detect, and so expand the range of calculable metallicities. However, as the line ratios are not a direct measurement of the metallicity, they must be benchmarked either to a subset of direct measurements where possible, or to theoretical models which can account for varying physical parameters in the gas cloud. 
Strong line metallicity calibrations based on theoretical models are each tied to their own set of models, such as the photoionization models  {\sc{cloudy}} \citep{Ferland1998}, or {\sc{mappings}} \citep{Groves2004a, Groves2004b}, which are often paired with a spectral synthesis model such as {\sc{starburst99}} \citep{Leitherer1999} underpinning the source of the stellar ionization field. These differences in model choices and in emission line ratios chosen as the tracers of metallicity will result in different numerical values of the resultant metallicity, often varying by 0.5 dex \citep{Kewley2008}. 

Metallicities calculated through different methods thus cannot be directly compared to each other due to differences in their zero points, resulting in the need for conversions between calibrations \citep{Kewley2008}. Such conversions permit comparisons between studies which, for reasons of preference or necessity, use different metallicity calibrations. For instance, instrument wavelength coverage and the redshift of the galaxy can conspire to put certain emission lines required for some metallicity calibrations out of range, limiting the choice of calibrations. 

The present work is by no means the first attempt to provide conversions between metallicity calibrations; \cite{Kewley2008} undertook this effort for a sample of 27,730 galaxies and 10 calibrations using the Sloan Digital Sky Survey \citep[SDSS;][]{York2000, Gunn2006} Data Release 4 \citep[DR4;][]{DR4}. Much more recently, \citet[][henceforth T21]{Teimoorinia2021} updated these conversions for the SDSS Data Release 7 \citep[DR7;][]{DR7}, including both a larger number of galaxies ($\sim61,000$) and a set of 11 calibrations, including more recently published metallicity calibrations, finding broadly that the increased statistics of the DR7 change the fits needed to match the data well. An alternate, Random Forest (RF) method was also found to reproduce conversions between calibrations equally well as a polynomial fit, without a need for a specific functional form or the need to break the degeneracy of double valued calibrations.   

Both the DR4 and the DR7 were part of the first two generations of the SDSS, where the spectrum of a galaxy was represented by a single 3 arcsecond fibre covering the central region of the galaxy \citep{York2000}. The current generation of the SDSS, SDSS-IV \citep{Blanton2017} includes an integral field spectroscopy (IFS) survey: Mapping Nearby Galaxies at Apache Point Observatory \cite[MaNGA;][]{Bundy2015, Law2015, Yan2016a, Yan2016b, Wake2017}, which aims to map 10,000 galaxies with bundles of 2 arcsecond fibres \citep{Smee2013}, ranging between 19 and 127 fibres per bundle \citep{Drory2015}. 
The MaNGA data set therefore represents a substantial increase in spectral data volume, even though it does not contain the sheer numbers of unique galaxies surveyed in the DR7. MaNGA also includes spectra at larger galactic radii than was possible with earlier data releases, and, as the survey is comprised of lower redshift galaxies, the spatial resolution is finer than that of the earlier SDSS survey generations, with a typical resolution element of 1.8 kpc at the typical redshift of $z \approx 0.03$ \citep{Law2016}. The Data Release 15 \cite[DR15; ][]{Aguado2019} is a cumulative, and the penultimate, data release of the MaNGA survey, comprising 4824 galaxies and over 9 million spectra. 

With the advent of large IFS surveys, many properties of galaxies, previously examined with a central fibre of the galaxy, are now being re-examined and modified with the advantage of both finer resolution and spectral coverage outside the galactic nucleus. In this context, we wish to re-examine the conversions between metallicity calibrations in the MaNGA DR15 sample. We aim to determine whether the spatially resolved data used here reveals a strong dependence on the spatial scale of the spectrum, or whether the conversions presented for integrated spectra are functional for IFS surveys. We therefore re-calculate conversions between commonly used metallicity calibrations for the DR15, assess the accuracy of these conversions, and compare these results to those of other surveys, both at lower and at higher resolutions. 

This work is organized as follows: 
in Section \ref{sec:calibrations}, we outline our quality assurance on strong emission lines and introduce the 11 different metallicity calibrations we use in the current work. 
In Section \ref{sec:conversions}, we explore the polynomial fitting procedure, additional restrictions imposed, and outline the conversions between calibrations. 
In Section \ref{sec:analysis}, we assess the scatter around the polynomial fits. We also compare the results of this work directly to surveys with different spectral resolutions, and assess these results for the influence of known observational biases. 
We conclude with a brief discussion in Section \ref{sec:discussion} and summary of our results in Section \ref{sec:conclusions}, and 
present a complete set of figures and tables in the supplementary material, available online. 
We assume cosmology of H$_0$=70 km s$^{-1}$ Mpc$^{-1}$, $\Omega_M$=0.3, and $\Omega_{\Lambda}$=0.7.

\section{Metallicity calculations}
\label{sec:calibrations}

\subsection{Sample selection \& Quality Control}
\label{sec:QA}

Emission line data are taken from the Pipe3D publicly processed MaNGA data \citep{Sanchez2016, Sanchez2016b}. Fluxes are then dust corrected, assuming an optically thick Balmer decrement of 2.85, for all spaxels where the S/N in the H$\alpha$/H$\beta$ ratio is $> 5.0$. Dust corrections in the target spaxels are undertaken with two extinction curves, one with a Small Magellanic Cloud (SMC)-like dust curve \citep{Pei1992} and one with a Milky Way (MW)-like curve \citep{ccm}, with characteristic values of E(B--V) of 0.123 and 0.096 respectively. The two dust correction curves result in slightly different final fluxes for the emission lines considered, which has the consequence of slightly altering the emission line ratios. For this reason, all of the analysis which follows is conducted for both the MW and the SMC dust correction curves independently. 

To ensure the validity of the strong emission lines required of the metallicity calibrations we consider for the remainder of this work, we also require classification on the AGN diagnostic diagram \citep[henceforth BPT]{Baldwin1981}. We require a S/N of  $> 5.0$ in the four emission lines required; \OIII$\lambda5007$, H$\alpha$, H$\beta$, and \NII$\lambda6584$. Each spaxel is classified according to its location on the diagram as AGN or star forming by the diagnostic line of \citet{Kauffmann2003}. If a given spaxel is considered star forming by  \citet{Kauffmann2003}, it remains in the sample. The BPT classification also introduces a difference in sample size between the MW \& SMC samples. While the dust correction does not affect the total number of classifiable spaxels, the choice of dust correction model alters the distribution of the spaxels across the BPT diagram. Out of a total of 3,063,241 classifiable spaxels, the SMC dust correction results in 2,332,924 (76.1 per cent) spaxels passing the \citet{Kauffmann2003} criteria. The MW dust correction results in 2,2297,702 (72.8 per cent) star forming spaxels by the same criteria.

We also impose an \ha EW cut. Following the work of \cite{Fernandes2011}, we impose an \ha EW threshold $>6 $ \AA~ to eliminate ionization regions not dominated by young star formation, but by aged stars or shocked gas \citep{Sanchez2018}. To prevent spuriously large \ha EW values from being included, we also limit the EW values to $<1000$ \AA. The inclusion of this limitation of 6 \AA~$<$ EW $< 1000$ \AA~does not substantially reduce the number of spaxels with metallicities remaining in the sample (SMC \& MW dust corrected fluxes lose $\sim$0.6 and $\sim$0.5 per cent of their sample, respectively).

Many of the metallicity calibrations require additional emission lines beyond the four required of the BPT diagram; on a per-calibration basis, we require S/N $> 5.0$ in any additional lines required by a given calibration, matching the S/N cut imposed on the BPT classification lines.

\subsection{Metallicity Calibrations}

For this work, we compare 11 different metallicity calibrations. This list includes the majority of the calibrations included in \citet{Kewley2008} and is supplemented by the inclusion of calibration methods which have appeared in the literature since the publication of that work in 2008. This set of 11 is identical to those included in \citet{Teimoorinia2021}. We include a brief summary of the original works, but refer the reader to each publication, where the calibration is described in full. 

\subsubsection{R$_{23}$-based calibrations}
The R$_{23}$ ratio is defined \citep{Pagel1979} as the sum of the \OII~and \OIII~lines divided by the H$\beta$ flux:

\begin{equation}
	\mathrm{R}_{23} = \frac{\text{\OII}\lambda\lambda3727+\text{\OIII}\lambda\lambda4959,5007}{\mathrm{H}\beta}
\end{equation}

R$_{23}$ is notoriously double-valued with chemical abundance, and theoretical models indicate that it additionally depends on ionization parameter \citep[e.g.,][]{Kewley2002}. 
The double valued nature of the dependence of R$_{23}$ on metallicity is due to two distinct regimes in the interstellar medium (ISM).  At low metallicity, R$_{23}$ correlates with O/H.  However, as metallicity increases, the ISM is able to cool more effectively, leading to an inverse correlation between R$_{23}$ and metallicity above a value of log(O/H)+12 $\sim$ 8.  Therefore, for a given value of R$_{23}$, there is both a high and a low metallicity solution.
Metallicity calibrations using this line ratio thus have two `branches': an upper branch (high metallicity), and a lower branch (low metallicity).
To choose between the degenerate upper and lower branch solutions, a second emission line ratio is required. 
Typically, this second emission line ratio involves the \NII$\lambda6584$ line, and for all the calibrations examined in the present work, the \NII/\OII~line ratio is employed\footnote{The N/O ratio depends on the relative abundances of these two metals. As nitrogen can be produced by more than one synthesis pathway, this ratio is not always a constant \citep{Belfiore2017}.}. 

We make use of four published calibrations of R$_{23}$ to a metallicity. While measuring the R$_{23}$ line ratio itself is straightforward, converting from this ratio into a metallicity requires the use of photoionization models, which each of these 4 published calibrations approach slightly differently.  These calibrations are: \citet[][hereafter Z94]{Z94}, which uses the average of 3 different theoretical models to convert R$_{23}$ into a metallicity. \citet[][M91]{m91} benchmarks the R$_{23}$ ratio to {\sc{cloudy}} photoionization models to calculate a metallicity. \citet[][KK04]{KK04} uses R$_{23}$ in combination with an iterative approach to estimate ionization parameter; this estimate of ionization parameter and emission line are converted to a metallicity via the theoretical model outlined in \citet[][KD02]{Kewley2002}: the photoionization code {\sc{mappings III}} \citep{Sutherland1993, Sutherland2013}, using both {\sc{starburst99}} and {\sc{pegase}} \citep{Fioc1997} stellar synthesis codes as a basis. Finally, the calibration of \citet{Kewley2002} as reformulated in \citet[][KE08]{Kewley2008}\footnote{Throughout this work we will refer to the calibration as KE08, and provide the citation for the publication in which it appeared.}. This reformulation uses the same theoretical model as KD02, which it therefore shares with KK04, but for lower branch values of R$_{23}$, an average of the lower branch M91 and KK04 solutions is used. 

\subsubsection{N2-based calibrations}

The second of the commonly used emission line ratios is N2, proposed by \cite{StorchiBergmann1994}, and which is defined as:

\begin{equation}
\text{N}2 =\frac{\text{\NII} \lambda6584}{\text{H}\alpha}
\end{equation}

 This line ratio has the advantages of being much simpler, requiring only two emission lines which are not widely separated in wavelength space (making them relatively insensitive to wavelength-dependent dust reddening), and not requiring a degeneracy breaking step, unlike R$_{23}$. However, \NII~is somewhat sensitive to ionization parameter \citep{Denicolo2002} and does have a tendency to saturate at high metallicities \citep{Baldwin1981, Kewley2002} as \NII~becomes an important coolant of the ISM. 

We make use of three calibrations which use the N2 line ratio; these line ratios are benchmarked empirically to a subset of data where ``direct'' (electron temperature) metallicities are possible to measure. \citet[][PP04]{PP04} converts a set of 137 \HII~regions which have either direct or detailed photoionization model based metallicities against the \NII/\ha line ratio. Similarly, \citet[][M13]{Marino2013} calibrates against a sample of 3423 \HII~regions from the CALIFA survey for which electron temperatures are calculable, and \citet[][C17]{Curti2017} provide N2 calibrations using stacks of $\sim$110,000 spectra from the SDSS DR7 to obtain ``direct'' metallicity estimates.

\subsubsection{O3N2-based calibrations}

O3N2, first presented in \cite{Alloin1979} and revisited by \cite{PP04} is defined as:

\begin{equation}
	\mathrm{O3N2}=\frac{\text{\OIII}\lambda5007/\mathrm{H}\beta}{\text{\NII}\lambda6584/ \mathrm{H}\alpha}
\end{equation}

The O3N2 line ratio, like N2, is designed to incorporate emission lines which are close in wavelength space, which is particularly convenient for high redshift spectral abundance measurements, where observing time is expensive, and was proposed as an alternative to the N2 calibrations, as the O3N2 ratio does not saturate at high metallicity. However, it is more sensitive to changes in ionization parameter than the simpler N2 calibration \citep{Kewley2019}.

 Each of the N2-based calibrations presented above is also paired with an alternate metallicity calibration based on the O3N2 line ratio (PP04, M13, and C17). These O3N2-based calibrations are empirically benchmarked to the same direct metallicity subsamples as their N2-based counterparts; we include the O3N2-based calibrations presented in PP04, M13, and C17.

\subsubsection{N2S2-based calibration}
Finally, we include \citet[][D16]{Dopita2016}, which uniquely among our set requires the sulfur doublet \SII~$\lambda\lambda6717,6731$. This calibration, like the R$_{23}$-based calibrations, is based upon photoionization models to convert the line ratio into a metallicity; in this case {\sc{mappings 5.0}} \citep{Sutherland2018}. This calibration requires \NII, \ha, and the \SII~lines; the polynomial form requires both the N2 ratio (\NII/H$\alpha$) and \NII/\SII. These lines are close in wavelength, so dust corrections should be minimal, and is designed to be insensitive to variations in both ionization parameter and pressure, including in their theoretical models both high ionization fields and high pressures, to account for these regions in the high redshift universe. 

\subsubsection{The final set of calibrations}
With the four R$_{23}$-based calibrations, three N2-based calibrations, three O3N2-based calibrations, and one N2S2-based calibration, our sample comprises 11 unique methods. Z94, M91, KK04, KE08, PP04 N2, PP04 O3N2 were included in \citet{Kewley2008}; we additionally include D16, M13 N2, M13 O3N2, C17 N2, and C17 O3N2.  
Each of these calibrations has a different range of validities, which are sometimes specified in terms of a 12+log(O/H) metallicity value and sometimes specified in terms of the line ratio on which they are based. Common to all of these calibrations is the need for the \NII$\lambda6584$ line.
A summary of all 11 calibrations, their required lines, and their range of validity, and whether or not they have an upper and lower branch is given in Table \ref{tab:ohlist}. 

\begin{table*}
\centering
\caption{List of the considered metallicity calibrations and their range of validity, and emission lines required. The KD02 metric only requires H$\beta$ and \OIII$\lambda5007$ for the lower branch.}
\label{tab:ohlist}
\begin{tabular}{lrlr}
\hline
  Metallicity calibration & Range of validity & Lines required & Upper branch dividing point\\
  \hline
    \citealt{Z94} (Z94) & log(\NII/\OII) $>$ -1.2 & \OII$\lambda$3727, \OIII$\lambda$5007 & Upper branch only:  \\
 & \& 12+ log(O/H) $>$ 8.35& H$\beta$, \NII$\lambda$6584& log(\NII/\OII) $>$ $-$1.2\\
  \hline
    \citealt{m91}  (M91)  & 7.24 $<$12+ log(O/H) $<$9.4 &  \OII$\lambda$3727, \OIII$\lambda$5007 & log(\NII/\OII) $>$ $-$1.2 \\
&& H$\beta$, \NII$\lambda$6584 & \\
  \hline
  \citealt{Kewley2002} (KD02) &  8.2 $<$12+ log(O/H) $<$ 9.4 &  \OII$\lambda$3727, \NII$\lambda$6584 & log(\NII/\OII) $>$ $-$1.2  \\
 \citealt{Kewley2008} (KE08) & & H$\beta$, \OIII$\lambda$5007  \\
  \hline
  \citealt{KK04}   (KK04)  & 7.6 $<$12+ log(O/H) $<$ 9.2&  \OII$\lambda$3727, \NII$\lambda$6584, & 12+ log(O/H)$>$8.4 \\
 && H$\beta$, \OIII$\lambda$5007 & log(\NII/\OII) $>$ $-$1.2 \\
  \hline
	  \citealt{Dopita2016} (D16)  &  7.76 $<$12+ log(O/H) $<$ 9.05 & \NII$\lambda$6584, H$\alpha$,  \SII~$\lambda\lambda$6717, 6731 & ...\\
    \hline
   \citealt{PP04} (PP04 N2) & -2.5 $<$ N2 $<$ -0.3 & \NII$\lambda$6584, H$\alpha$ & ...\\
  \hline
    \citealt{PP04} (PP04 O3N2)  & O3N2 $<$ 2.0 & \NII$\lambda$6584, H$\alpha$, \OIII$\lambda$5007, H$\beta$ & ...  \\
  \hline
    \citealt{Marino2013}  (M13 N2) & -1.6 $<$ N2 $<$ -0.2 &\NII$\lambda$6584, H$\alpha$ & ... \\
  \hline
  \citealt{Marino2013} (M13 O3N2)  & -1.1 $<$O3N2$<$ 1.7& \NII$\lambda$6584, H$\alpha$, \OIII$\lambda$5007, H$\beta$ & ...\\
  \hline
     \citealt{Curti2017}   (C17 N2)  & 7.6 $<$12+ log(O/H) $<$ 8.85 &\NII$\lambda$6584, H$\alpha$ & ... \\
  \hline
     \citealt{Curti2017}  (C17 O3N2) & 7.6 $<$12+ log(O/H) $<$ 8.85& \NII$\lambda$6584, H$\alpha$, \OIII$\lambda$5007, H$\beta$ & ...\\
  \hline
\end{tabular}
\end{table*}

\begin{table}
	\centering
	\caption{Total number of metallicity values per calibration, for both the SMC and MW dust correction models, after S/N cuts, BPT classifications, and including an \ha EW cut. MW-dust corrected spaxels are fewer in number due to a reduced fraction of them passing the BPT classification.}
	\label{tab:oh_nums}
	\begin{tabular}{lcr}
	\hline
		Calibration & SMC spaxels: & MW spaxels:\\
		\hline
		Z94 & 1,047,232 & 858,244 \\ 
		M91 & 1,070,812  & 962,022 \\ 
		KK04 & 1,065,094 & 989,498 \\ 
		KE08 &1,069,318  & 1,027,191 \\ 
		D16 & 1,012,424  & 978,738  \\ 
		PP04 N2 & 1,091,383  & 1,055,164  \\ 
		M13 N2 & 1,092,348 & 1,055,271  \\ 
		C17 N2  &  1,081,521  & 1,051,017  \\ 
		PP04 O3N2 & 1,090,897  & 1,053,272 \\
		M13 O3N2 & 1,081,897  & 1,042,398  \\ 
		C17 O3N2 & 1,092,428  & 1,055,361  \\ 
		
		\hline
	\end{tabular}
\end{table}

For each calibration, we follow the prescriptions given in their respective papers to calculate metallicities for all possible MaNGA spaxels. The final counts of the number of spaxels with metallicities for each of the calibrations, for both the SMC and MW dust correction curves, is given in Table \ref{tab:oh_nums}. 

\section{Conversions between calibrations}
\label{sec:conversions}

For each pairing of calibrations, we select all spaxels with both metallicities present, given the dust correction method choice; either SMC or MW. As different calibrations are valid over different metallicity ranges, and may require a larger or smaller number of significant emission lines, the number of spaxels with calculated metallicities in common varies by several tens of thousands for each pairing, though typically remains of order 1 million spaxels. 
The number of spaxels included in each pairing is listed in Appendix Table \ref{tab:smc_values}, and in full in the supplementary material.

\subsection{Polynomial fitting procedure}
\label{sec:polyfit}
The correlations between calibrations vary substantially both in scatter and in functional form. 
Many of these calibrations show rather complex features in their correlations; for others the correlations are nearly linear.
Our polynomial fitting procedure reflects the complexity of the data. A simple ordinary least squares (OLS) fit to the full data set fails to accurately trace the median of the distribution. In this data set, there is often significant skew around the median $y$ value at fixed $x$ metallicity value. For calibrations that have large scatter, this skewness typically trends low at high $x$ metallicity, and high at low $x$ metallicity. An OLS fitting routine presumes some level of gaussianity to the data distribution, which is not present in this particular data. Therefore, in an attempt to reduce the overall distance from the fitted curve, the OLS tends to result in a polynomial fit which is flatter than desired, often missing the median ridge entirely. 

We approach the polynomial fit with the specific aim of reducing skew between the polynomial fit and the upper and lower envelopes of the data. We thus identify, for each of 100 bins in the horizontal axis, the 99.7th, 98.8th, 95.5th, 86.6th, and 68.3rd percentile of the data, corresponding to $3\sigma, 2.5 \sigma, 2 \sigma, 1.5 \sigma,$ and $1 \sigma$ vertical scatter, which provides contours around the data distribution. Repeating this exercise with 1 minus the above percentiles produces the lower bound. We then select all the individual data points within each of these contours and perform a polynomial fit\footnote{numpy.polynomial.polynomial.Polynomial.fit} to the raw data points within these bounds. If there are fewer than 10 spaxels in a horizontal bin, it is omitted. This omission is in play for the polynomial fit only for the very upper and lower $x$ values of the sample. We assess the performance of this fit by finding the median offset between the selected contour boundaries and the polynomial fit in both the positive and negative direction. Skew is assessed by the difference in these median offsets. If the negative contour were closer to the polynomial fit, then the polynomial fit is systematically underpredicting the correlation. We iterate through all percentile bounds, and select as the best polynomial fit whichever has the smallest residual offsets between the upper and lower percentile bounds. This is usually (but not required to be) the 1$\sigma$ contour, which most closely traces the bulk of the metallicity values. The selected best-fit polynomials are presented in Appendix Table \ref{tab:smc_values}, and in full in the supplementary material. 

For $x$ calibrations with upper and lower branches, this process is repeated for both branches, with a separate percentile finding and polynomial fit process. Given the smaller data range in the lower branch, a second order polynomial is used for the fit, and the number of horizontal bins is reduced to 30. All other processes are kept the same. 

\begin{figure*}
\includegraphics[width=.32\textwidth]{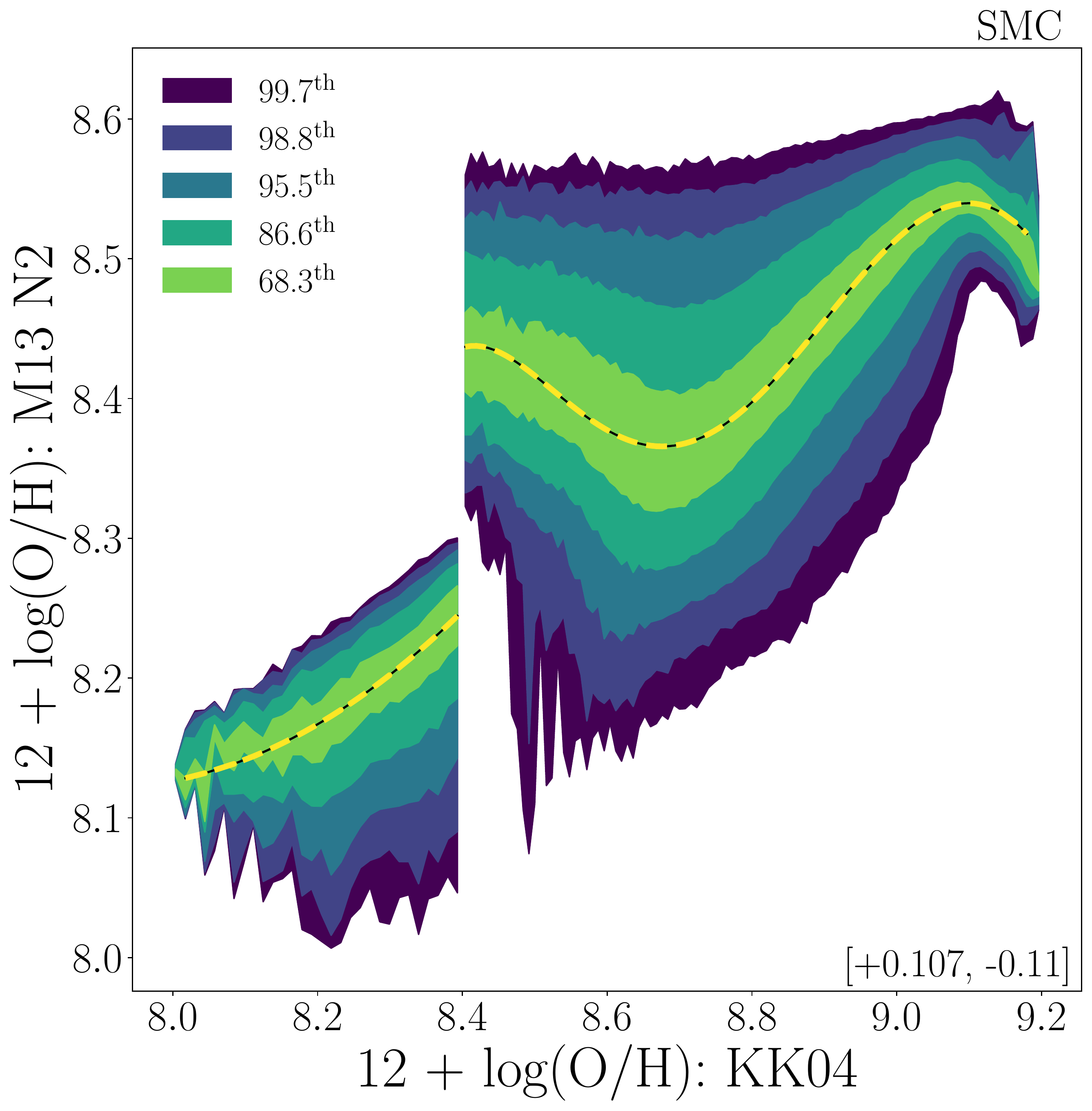}
  \includegraphics[width=.326\textwidth]{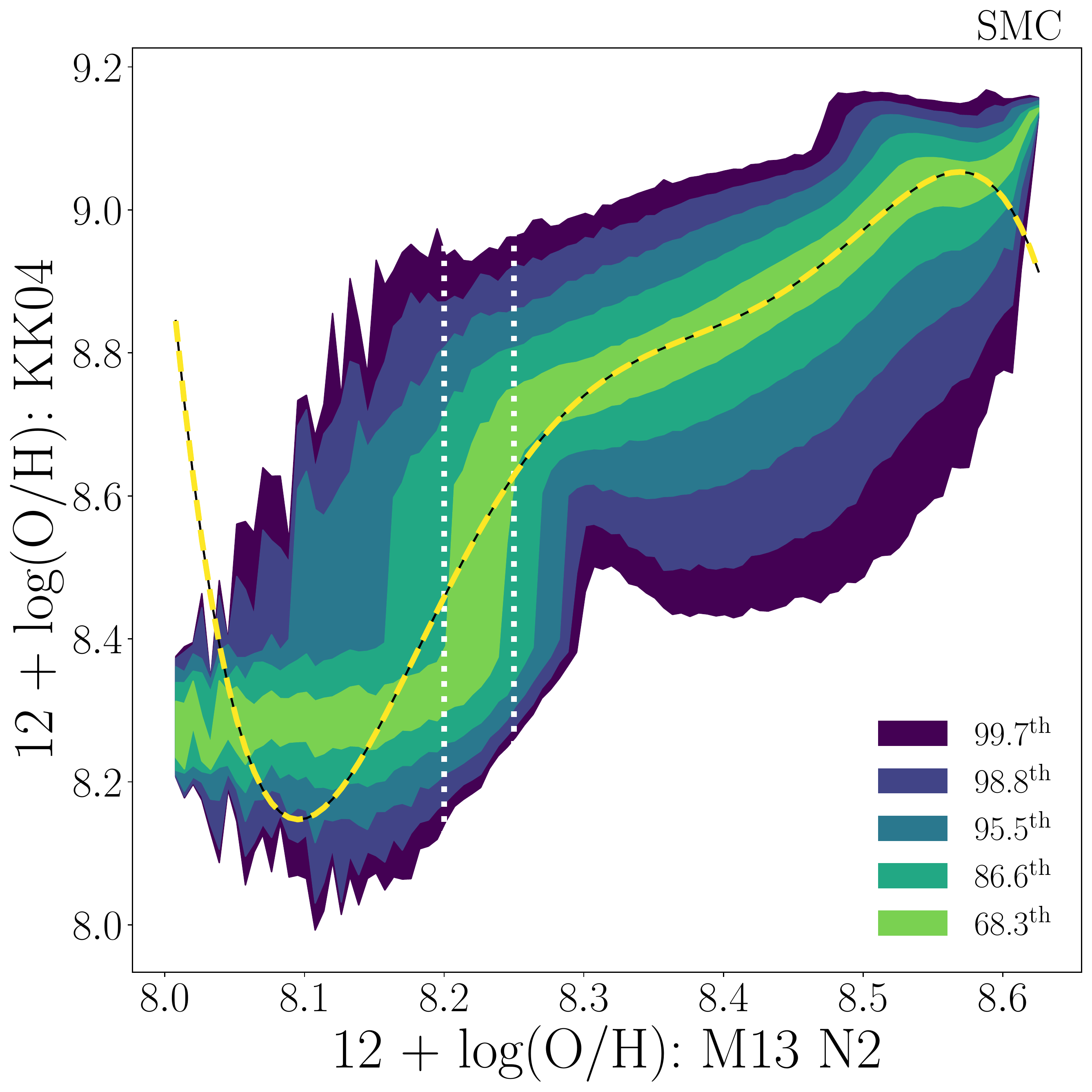}
   \includegraphics[width=.32\textwidth]{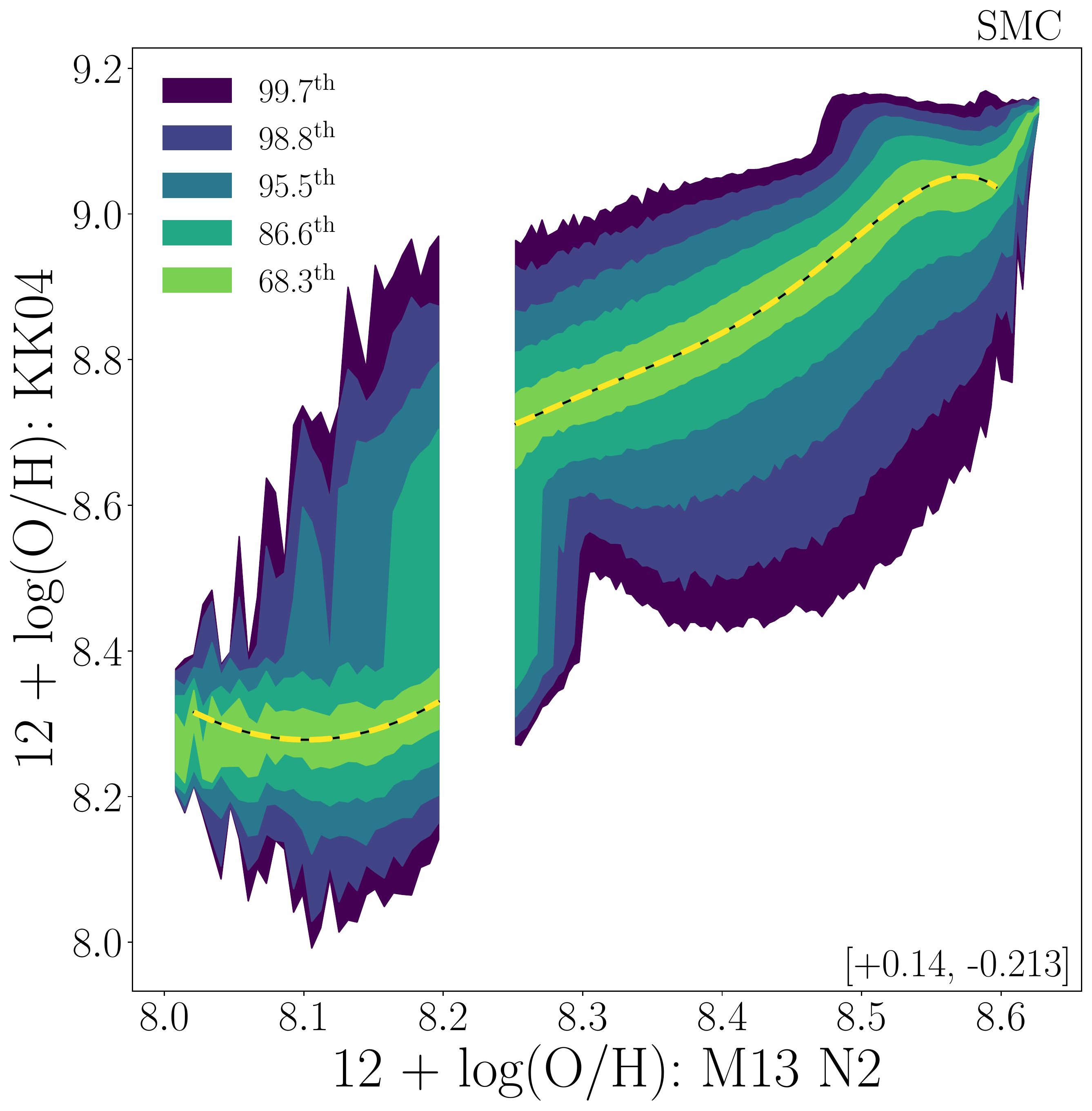}
  \caption{Left hand panel: Conversion from KK04, a conversion with an upper/lower branch division at log (O/H) + 12 = 8.4. A yellow-black dashed line shows the results of a 5th order polynomial for this data. Upper and lower branches are fit separately. The legend indicates the percentile ranges by colour, which range from 99.7th percentile in dark purple, to 68.3rd percentile in light green. The upper metallicity end has been truncated to maintain the polynomial only where it is within the 1$\sigma$ contours. Center panel: The axis-reversal of the left hand panel. The conversion from M13 N2, a calibration with only a single branch, into KK04, a calibration with upper and lower branches. A yellow-black dotted line shows the results of a poorly fitting 5th order polynomial for this data, without the application of the 1$\sigma$ contour clipping. Plotted in vertical white dotted lines are a manual superposition of 8.2 \& 8.25 to flag the most unreliable portion of this data. Right panel: The results of the additional quality control imposed. Upper and lower regimes are fit separately, and a range limit has been imposed on the higher metallicity end to prevent the polynomial from departing the 68.3rd percentile range.\label{fig:poorfit}}
\end{figure*}

\begin{figure*}
  \includegraphics[width='7in]{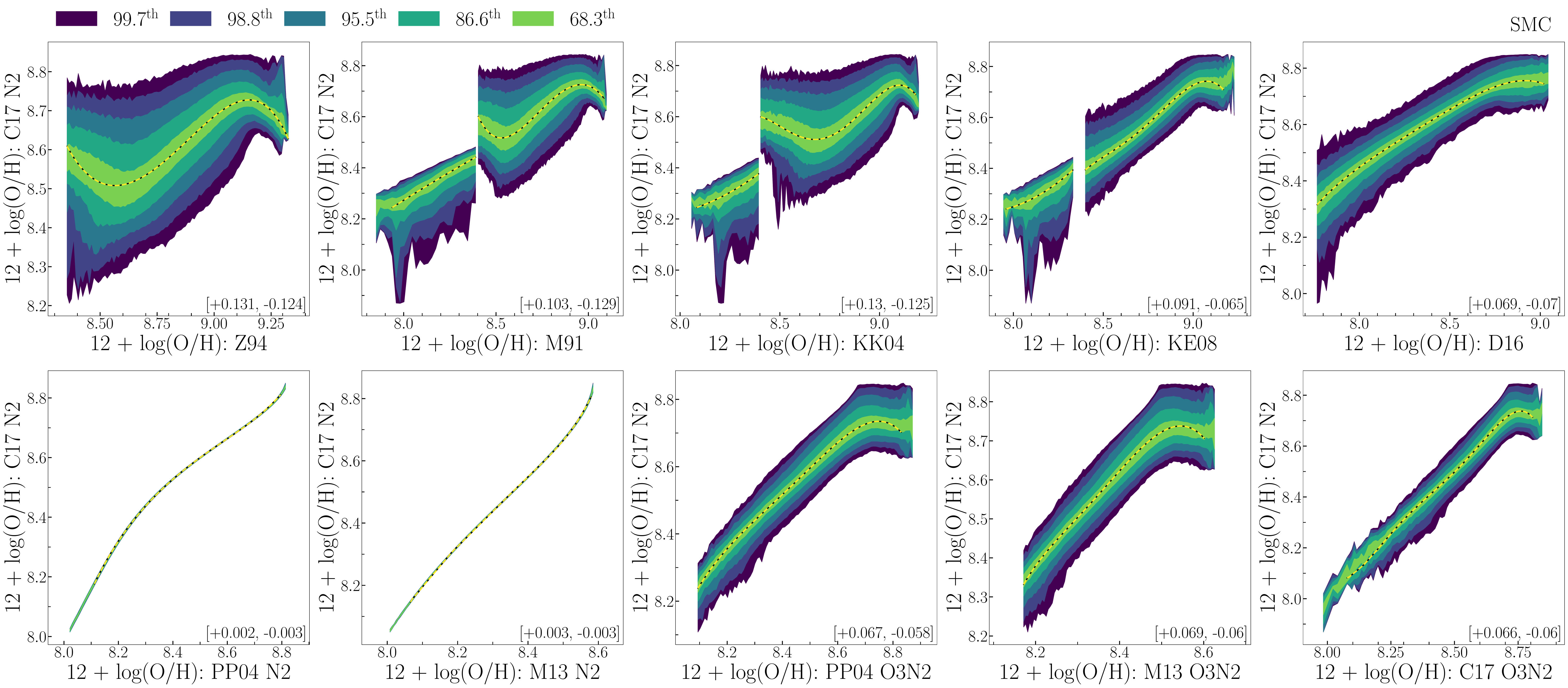}
  \caption{Correlations between all metallicities and C17 N2, assuming an SMC dust correction. For each calibration pair, we plot the area which encloses the 99.7th, 98.8th, 95.5th, 86.6th, and 68.3rd percentiles of the data, in dark blue to pale green respectively. Superimposed on this data range is the best fit polynomial plotted in a yellow-black dashed line. In the bottom right corner we indicate the median separation of the 2$\sigma$ (95.5th) contour from the line of best fit, in dex. This set of conversions typifies the range of correlation morphologies present across the data set. The complete set of correlations is available on line.}
  \label{fig:intoC17N2}
\end{figure*}

\subsubsection{Upper and lower bounds on the conversions}
While our polynomial fits are performed on the full $x$ range of the data (excluding bins with less than 10 spaxels), there are conversions where the polynomial does not well match the data at the upper or lower limits. An extreme example of this issue is seen at both the high \& low metallicity end of the centre panel in Fig. \ref{fig:poorfit}, where the polynomial deviates subtantially outside the range of the data.  In these cases, we truncate the range of validity of our calibration conversions. This is usually a small truncation, both in terms of metallicity range and in the fraction of spaxels which are excluded as a result; the typical total reduction in range is 0.04 dex. To implement this limitation in a consistent way, we find the $x$ value which corresponds to the polynomial being out of the bounds of the 68.3rd percentile ($1\sigma$) region, either in the positive or negative direction. If the polynomial fit remains within the region that encompasses the data across all $x$ values, no further cuts are implemented. However, in cases where the polynomial fit deviates strongly from the bulk of the data, we are able to limit the functional range of the calibration conversion. The second from right and right hand panels of Figure \ref{fig:intoC17N2} (both upper and lower rows) show the results of this truncation at the high $x$ metallicity end.  The range of validity presented in Table \ref{tab:smc_values}, and the polynomial fits shown in all figures (excepting the exemplar centre panel of Figure \ref{fig:poorfit}) have undergone this additional criterion. 

There are 110 possible pairs between metallicity calibrations, including the two directions to convert from and to. With approximately a million spaxels in each pairing, plotting even a density histogram to display all of them simultaneously is prohibitively memory-intensive, and even a subset of the data is rapidly impractical. Therefore, for presentation purposes, we simply show the edges of our percentile contours identified for our fitting procedure. The results of this polynomial fit method are shown in Figure \ref{fig:intoC17N2}, which shows, from dark blue to green, the regions enclosing the percentile ranges from 99.7th to 68.3rd, respectively, for all metallicities paired with C17 N2 as the $y$ axis. C17 N2 contains most of the data pathologies encountered in the full data set. For the results across all combinations of calibrations, see the on line supplementary material for SMC \& MW dust correction curve figures\footnote{Due to the large data volumes treated in this work, we have opted to present the results for the SMC dust correction curves. The SMC curve is consistent with previously published work, and does not substantially differ from the MW dust correction curve.}.
The 5th order polynomial line of best fit is shown as a yellow/black dashed line in all panels.

\subsubsection{The choice of 5th order polynomials}
Previous works (\citealt{Kewley2008}; T21) have used 3rd order polynomials. If 5 orders are not required, the polynomial fit method we use will return very small coeffecients for the higher order terms. However, to directly assess the relative performance of the 5th order vs a 3rd order polynomial fit, we have tested the results of both 5th order and 3rd order polynomial fits. In general, the two fits are comparable, with a typical difference in final range (the $x$ range where the polynomial remains within the 1$\sigma$ contour) of only 0.008 dex wider in the 5th order polynomial case. However, the 5th order polynomial can, in some cases, extend the valid range of the conversion by 0.34 dex, while the 3rd order maximally extends the range by 0.07 dex. Given the considerable advantage in extending the data range for some conversions, and that neglecting the 3rd order option is not a major detriment for any conversion, we conclude that the 5th order fit is preferable. We also test whether re-fitting the polynomial after identifying the valid $x$ bounds to the fit improves the quality of the fit, and we find that this does not reduce skew in the fit, with median changes to the skew in the upper and lower contours of order 10$^{-4}$ dex. 

We also test, using the data from \citet{Teimoorinia2021}, that a 5th order polynomial is able to match the 3rd order polynomial reasonably well when fitting to the same data, and find that across the bulk of the data range, 3rd and 5th order polynomials are in near perfect agreement. We also verify that the percentile-based polynomial fitting routine (either 3rd or 5th order) can reproduce an OLS fit for the DR7 data set used by \citet{Teimoorinia2021}. At the very upper and lower ends of the data range, 5th order and 3rd order solutions do deviate from each other, but inspecting the data, the 5th order polynomial is again more faithfully tracing the data itself. We conclude that the 5th order polynomials are preferable and should not be strongly affected by overfitting issues. 

\subsection{Converting to calibrations with upper \& lower branches}
\label{sec:doublevalued}

The choice of $x$ and $y$ metallicities in Figure \ref{fig:intoC17N2} is completely arbitrary, and indeed for a complete set of conversions these figures should also be generated with their axes reversed. 
It is a coincedental feature of our choice of calibration ordering that calibrations with a lower branch (all R$_{23}$-based calibrations except Z94, which only has the upper branch) are preferentially on the horizontal axis. Plotting in the other direction reveals that converting from calibrations without upper/lower branches into calibrations with two branches results in a segment of double-valued metallicities, and extremely large scatter for metallicities which are found near the transition from upper to lower branch of the $y$ metallicity. The left and centre panels of Figure \ref{fig:poorfit} show an example of the dramatic change triggered by an axis reversal, with the left hand panel of Figure \ref{fig:poorfit} showing the calibration with two branches on the $x$ axis, and in the centre panel of Figure \ref{fig:poorfit}, that same calibration on the $y$ axis. 

The sharp vertical turn in the centre panel of Figure \ref{fig:poorfit} just above M13 N2 values of log(O/H)+12 = 8.2 is blurring across the upper and lower branches of KK04, which appears at log(O/H) + 12 = 8.4, more clearly visible in the left hand panel. In this calibration pairing, and generally across any such single-branched into a double branched calibration pairing, this vertical turn spans a vertical range of $\sim$0.5 dex, but a horizontal range of only $\sim 0.05$ dex. With KK04 on the vertical axis, a fifth order polynomial does not perform well to this contour; the results of such an attempt are shown in a black \& yellow dashed line in the centre panel of Figure \ref{fig:poorfit}, without the application of the 1$\sigma$ contour clipping. Increasing the degree of the polynomial fit also does not improve the quality of the fit. 

As the purpose of this work is to provide practical conversions between calibrations, this region of extremely broad scatter is problematic. However, this upturned region (bounded manually in vertical white dashed lines in the centre panel of Fig \ref{fig:poorfit} at 8.2 \& 8.24) is relatively narrow. We therefore opt to exclude this region from the range of valid metallicity conversions. An attempt to convert a M13 N2 calibration into KK04 in this metallicity range is functionally meaningless; even the 1$\sigma$ contour spans a wide range ($>$0.4 dex) of metallicity values. These exclusions do not result in the loss of a substantial fraction of valid spaxels; for the SMC (MW) dust correction method, the loss is 0.97 -- 1.7 (0.5  -- 1.6) per cent of the total data set. 

\begin{figure*}
  \includegraphics[width='7in]{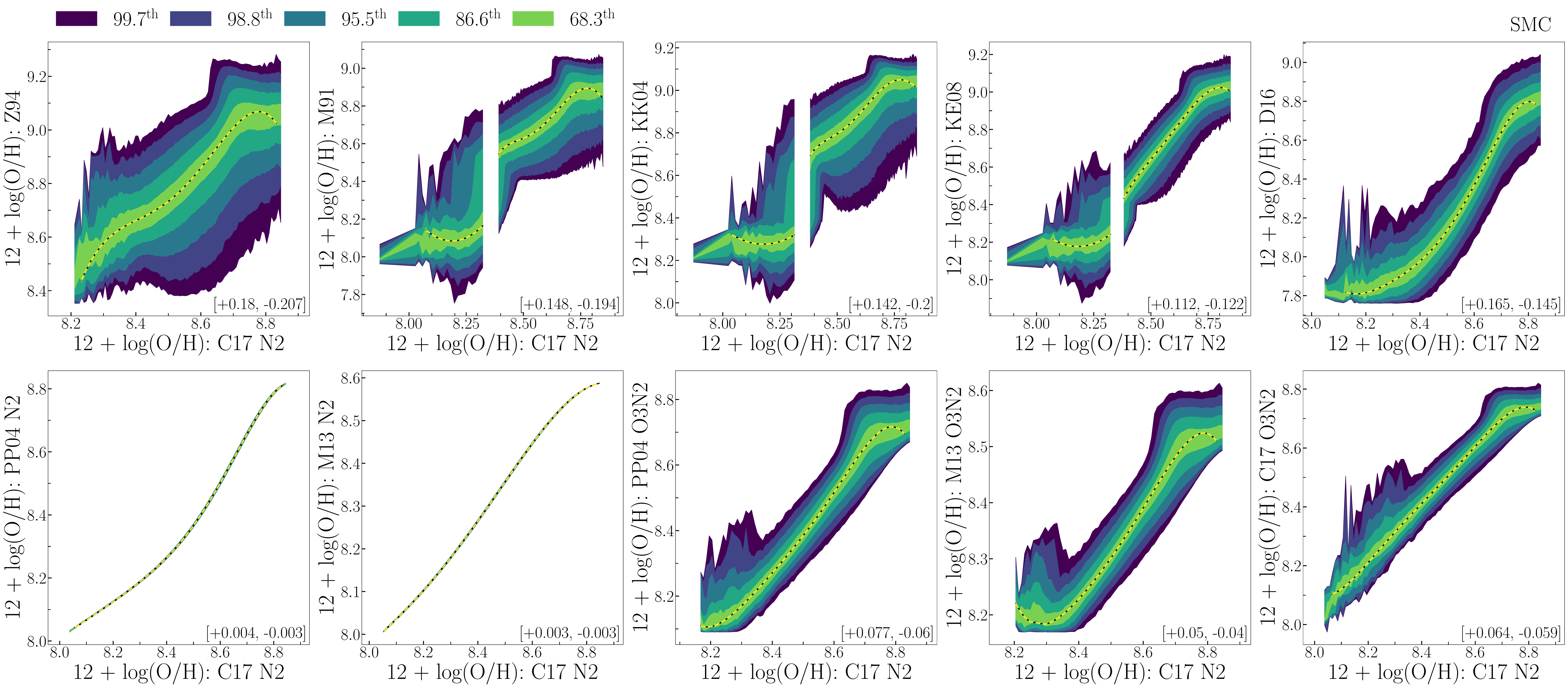}
  \caption{Conversions from C17 N2 into all other metallicities. This Figure is the same as Figure \ref{fig:intoC17N2} but with each panel's axes reversed. This set of conversions typifies the range of correlation morphologies present across the data set. Horizontal bins which have fewer than 10 spaxels are excluded from this visualization. The complete set of correlations is available on line.}
  \label{fig:fromC17N2}
\end{figure*}

This exclusion region, which we have manually set for affected conversions, imposes a divide in the validity of a metallicity calibration which would not otherwise have one, when converting into a calibration which is double-valued. However, since such divisions in the calibrations are common among our metallicity calibrations with upper and lower branches, we continue by applying the same method as is used for upper and lower branch calibrations, applying a 5th order polynomial to the upper component, and a 2nd order polynomial to the lower component. With the exclusion region set, the polynomial fits are able to match the centroid of the data as expected. The right hand panel of Figure \ref{fig:poorfit} shows the results of the polynomial fitting procedure once the exclusion region has been set; the black and yellow dashed line which marks the polynomial now follows the 1$\sigma$ region much more faithfully. The inverse axis ordering of Figure \ref{fig:intoC17N2} is shown in Figure \ref{fig:fromC17N2}, with this additional cut imposed. The vertical blank spaces midway through the data range in the upper, central three panels are the results of our masking procedure, and not a true gap in the data.  

Calibrations affected by this issue are any of: C17 N2, C17 O3N2, M13 N2, M13 O3N2, PP04, \& D16 into any of the R$_{23}$-based calibrations M91, KK04, \& KE08, for a total of 18 conversions between calibrations. We note that the dust calibration does make a slight difference in the positioning of this exclusion region, and so we have imposed separate exclusion regions depending on dust correction. The positions of these regions are included in our tabulation of the range of validity in Table \ref{tab:smc_values}.

\subsection{Residuals from the polynomial fit}
\begin{figure*}
  \includegraphics[width='7in]{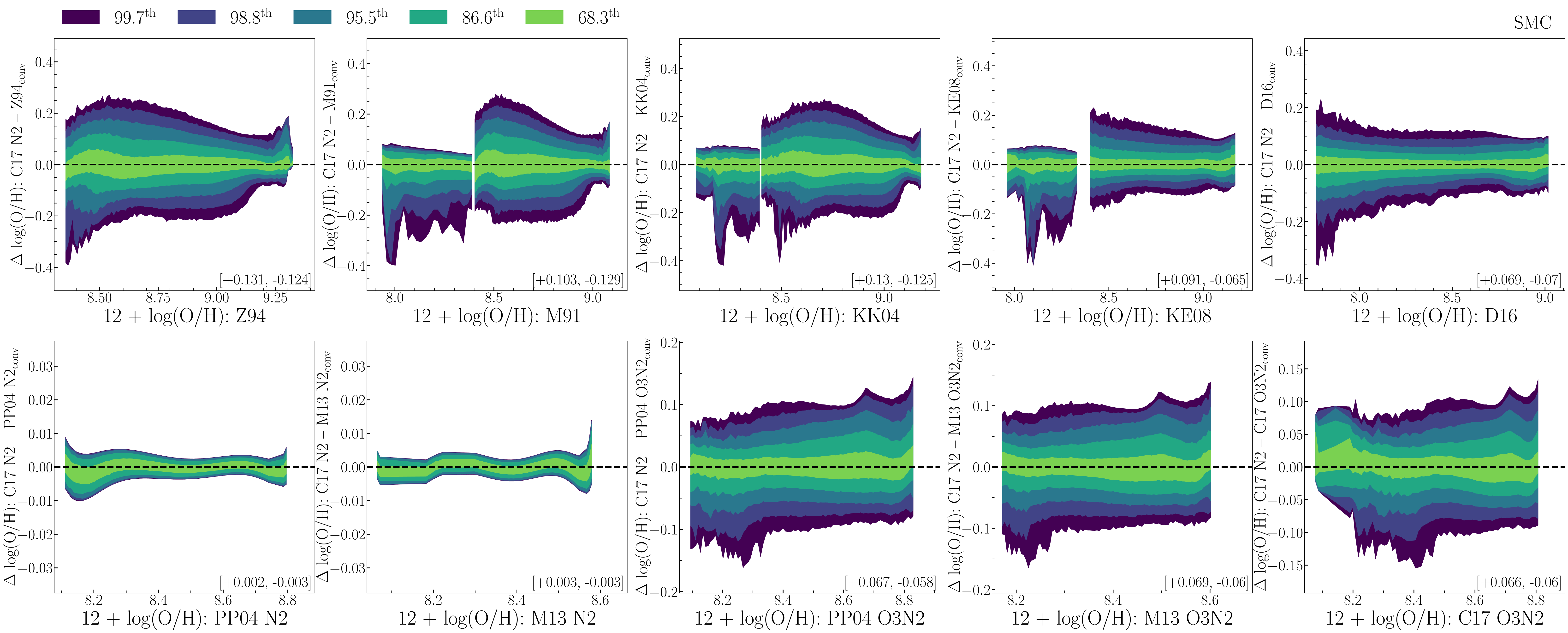}
  \caption{Residuals from the conversion for metallicity calibrations into C17 N2. The horizontal dashed line indicates the normalized position of the polynomial line of best fit. Each solid line is the contour enclosing 99.7, 98.8, 95.5, 86.6, and 68.3 per cent of the data, from purple to green respectively. Axes are symmetric around zero, but the scaling is distinct for each panel due to the dynamic range across the 10 panels. The median values of the upper and lower 2$\sigma$ contour is presented in the bottom right corner of each panel. The complete set of correlations is available on line.}
  \label{fig:intoC17N2_resid}
\end{figure*}

To visually represent the quality of fit and the residuals of the method outlined thus far, we show in Figure \ref{fig:intoC17N2_resid} 
the results of subtracting each metallicity value from the polynomial line of best fit. This figure is an exact analogue to Figure \ref{fig:intoC17N2}. We plot the contours as percentiles of the full distribution; upper and lower bounds of the 99.7th, 98.8th, 95.5th, 86.6th, and 68.3th percentiles in the positive and negative directions are plotted. Vertical axes are chosen such that they are symmetric around the zero line, which represents the normalized position of the polynomial best fit curve. With few exceptions, the residuals are broadly symmetric around zero, though we note that the scatter often has some variation as a function of the $x$ metallicity. 
For the majority of the calibration pairings, the contours are largely consistent across the valid range of the $x$ metallicity, and so we can expect that a median value is a reasonable approximation of the typical scatter. 

As such, we report the median value for upper and lower bounds of the 2$\sigma$ (95.5th contour) in the bottom right corner of each panel of Figures \ref{fig:intoC17N2} -- \ref{fig:intoC17N2_resid}. These medians are the median values of the 100 (or 130, for calibrations with upper \& lower branches) points denoting our contours, and correspond to the median values of the third (middle) contour in Figure \ref{fig:intoC17N2_resid}. Many of these median bounds in the positive and negative directions are within 0.01 dex of each other, indicating that we have indeed reduced skew in the fitting procedure. In Table \ref{tab:smc_values}, we list the median vertical separation from the best fit polynomial in the positive and negative directions for all pairings. A full set of residual figures is available in the online supplementary material. 

\begin{figure}
  \includegraphics[width=\columnwidth]{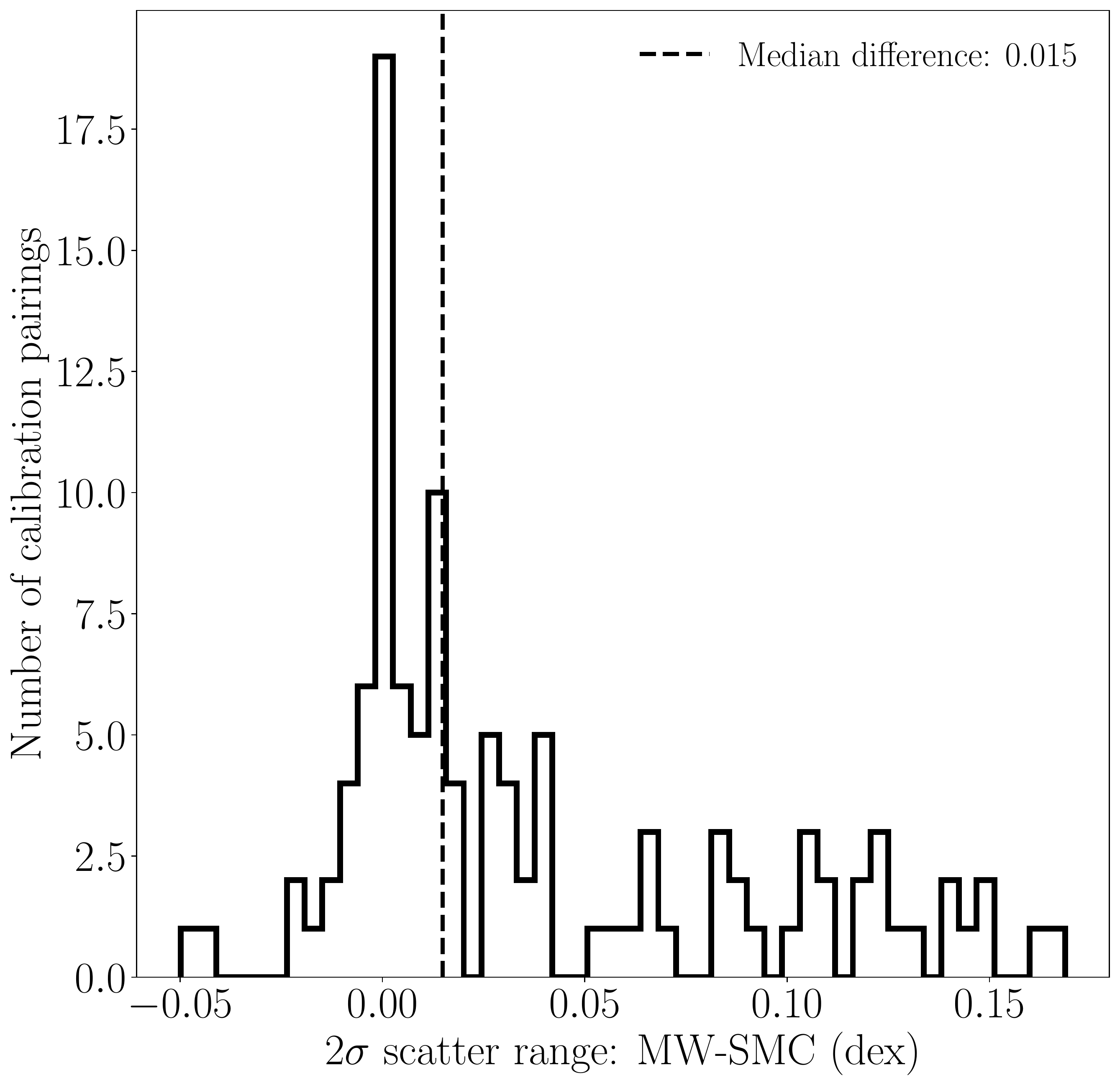}
  \caption{The change in 2$\sigma$ range, for the MW dust correction, vs the SMC dust correction, for each pairing. We find the upper - lower bounds of the 2$\sigma$ contour, subtract them to find the total range, and then subtract the SMC range from the MW range. In general, the MW dust correction curves result in slightly larger scatter, though the median difference between dust correction models is only 0.015 dex above zero. }
  \label{fig:dustcorr_scatter}
\end{figure}

The residuals traced by the 95.5th percentile are slightly larger for the MW dust curve corrections for the majority of the calibration pairings when compared to the SMC dust correction curve, but generally this difference is quite small (the median change in the range between upper and lower 2$\sigma$ contours is 0.015 dex), illustrated in Figure \ref{fig:dustcorr_scatter}.  Exceptions are present for the D16 \& Z94, M91, and KK04 pairings; the upper medians increase their offset by 0.1 dex when switching from SMC to MW curves, and there is a maximal change in the range spanned by the 2$\sigma$ contours of 0.17 dex. As with the previous figures, a complete, parallel set of figures for the MW dust correction curve is available as supplementary material. 

\section{Analysis}
\label{sec:analysis}

With all quality assurance mechanisms in place, and robust polynomial fits for conversions between all combinations of metallicity calibrations, we now examine these results in more detail. It is immediately apparent from Figure \ref{fig:intoC17N2} that there is not a single typical level of scatter for all conversions between metallicity calibrations. For conversions into C17 N2 alone, the level of 2$\sigma$ scatter varies from $\pm 0.003$ dex to +0.14/--0.124 dex. We therefore investigate possible sources of scatter in these conversions. We check for dependencies on required emission lines, and use comparison data sets to assess these conversions for observational biases. 

\subsection{Trends in the residual scatter}
\label{sec:scatter_trends}

There are six pairings which result in effectively zero scatter between the calibrations. Two of these are visible in Figures \ref{fig:intoC17N2} \& \ref{fig:fromC17N2}, in the lower left, and lower second from left panels of both Figures. The full set of these are between PP04 N2, M13 N2, \& C17 N2,  and between PP04 O3N2, M13 O3N2, \& C17 O3N2. Each set of these calibrations relies exclusively on the exact same emission line ratios, and all three, in each case, apply a polynomial calibration to either the N2 ratio or the O3N2 ratio. As the emission lines are identical for the same spaxels, we should expect zero scatter around the conversion between these calibrations. The scatter which does appear ($2\sigma$ scatter for these conversions is of order $\pm$0.003 dex) is due to a combination of the binning procedure used to determine percentile bounds, and our polynomial fit not providing an exact analytic solution between these polynomial calibrations. 

At the other extreme, certain calibrations appear to have particularly large scatter. The Z94, M91, and KK04 calibrations, while smoothly converting between each other, demonstrate extremely large scatters when converting into other calibrations. Most notably, the scatter for all three of these calibrations is large when converting into the D16 calibration. The only emission line in common between the D16 calibration and the Z94, M91, and KK04 calibrations is the \NII$\lambda$6584 line. We thus propose that the scatter in the conversions is at least partially dependent on the overlap in the strong emission lines required by each calibration. 

\subsubsection{Differences in required emission lines}
\label{sec:DR15_elines}
 To assess the dependence the scatter has on the overlap between the emission lines in a more quantitative fashion, we select each calibration pairing, and for each $x$ calibration, determine both which emission lines it has in common with the $y$ metallicity it is being converted into, and which emission lines are not considered by the $y$ metallicity. There are ten unique sets of emission line exclusions and inclusions between all 110 calibration pairs, three of which represent complete overlap in the emission lines required. For each unique set of emission line inclusions and exclusions, we collect the mean of the positive and the absolute value of the negative 2$\sigma$ residuals, as reported in the bottom corner of Figure \ref{fig:intoC17N2_resid}, for all calibrations which match. We then sort the emission line overlap by their median 2$\sigma$ residuals and present this sorting of the data in Table \ref{tab:eline_sort}. 

\begin{table}
	\centering
	\caption{In ascending typical 2$\sigma$ scatter, we present the emission line permutations between calibrations. For each set of calibrations which match the inclusions/exclusions, we find the typical offset of the 2$\sigma$ contour (the median absolute value of the 2$\sigma$ residuals) from our polynomial fit. We also include the smallest and largest $2\sigma$ residuals for each set. The median (and range) of the 2$\sigma$ scatter is smallest for all calibrations which have full overlap in their emission line requirements: the top row includes all of the O3N2-based calibrations. }
	\label{tab:eline_sort}
\begin{tabular}{r|lccc}
\hline
Lines excluded & Lines in overlap & \multicolumn{3}{|c||}{2$\sigma$ scatter (dex)} \\ 
($x$ but not $y$) & ($x$ \& $y$) & $\mu_{1/2}$ & min  & max  \\ 
\hline
Full overlap  &  \OIII, H$\beta$, H$\alpha$, \NII & 0.0024 & 0.0021 & 0.0036 \\
Full overlap  &  H$\alpha$, \NII  & 0.0543 & 0.002 & 0.1549 \\
Full overlap  &  \OII, \OIII,   H$\beta$, \NII & 0.0581 & 0.0135 & 0.1419 \\
\OIII, H$\beta$   &  H$\alpha$, \NII  & 0.064 & 0.0477 & 0.1994 \\
\SII   &  H$\alpha$, \NII & 0.0762 & 0.0522 & 0.1126 \\
\OII   &  \OIII, H$\beta$, \NII  & 0.091 & 0.0606 & 0.14 \\
\OII, \OIII, H$\beta$  &  \NII  & 0.128 & 0.0671 & 0.2874 \\
H$\alpha$   &  \OIII, H$\beta$, \NII  & 0.1419 & 0.1158 & 0.1663 \\
H$\alpha$, \SII   &  \NII   & 0.1564 & 0.1358 & 0.1863 \\
H$\alpha$   &  \NII & 0.1724 & 0.117 & 0.2023 \\
\hline
\end{tabular}
\end{table}

Table \ref{tab:eline_sort} shows the unique excluded and included emission lines, the typical (median) 2$\sigma$ scatter for all metallicity pairings with that set of emission line inclusions/exclusions, along with the minimum and maximum values of the 2$\sigma$ scatters. The \NII~line is in common between all pairings, and is always in overlap. 
The three pairings with complete overlap also are the pairings with the smallest median 2$\sigma$ scatter, and the conversions between O3N2-based calibrations are represented by the top row with a maximum $2\sigma$ scatter of 0.0036 dex. 
The largest typical scatter is found for calibrations which are being converted into a metallicity with no lines in common beyond the \NII~line. Calibrations which rely on the N2 ratio, or the D16 calibration which relies on H$\alpha$, \SII~and \NII~thereby perform most poorly when converting into the subset of calibrations which do not have \ha or \SII~line requirements respectively. Calibration pairs where the $x$ calibration requires the \ha line, and the $y$ calibration does not, also seems to correlate with generally higher levels of scatter.

Conversions from R$_{23}$-based calibrations, which require \OII, \OIII, H$\beta$, and \NII, can convert with minimal scatter (the minimum 2$\sigma$ scatter is $\sim0.013$ dex); but these calibrations can also produce large scatter, even to other R$_{23}$-based calibrations. The largest 2$\sigma$ scatter between R$_{23}$-based calibrations is 0.14 dex. This large scatter between two R$_{23}$-based calibrations (KK04 $\rightarrow$ KE08) indicates that the emission lines required are not the only controlling factor for the level of scatter. In the case of the R$_{23}$-based calibrations, the other variable is likely to be the theoretical models to which the R$_{23}$ line ratio is benchmarked. 

\begin{figure*}
  \includegraphics[width=1.3\columnwidth]{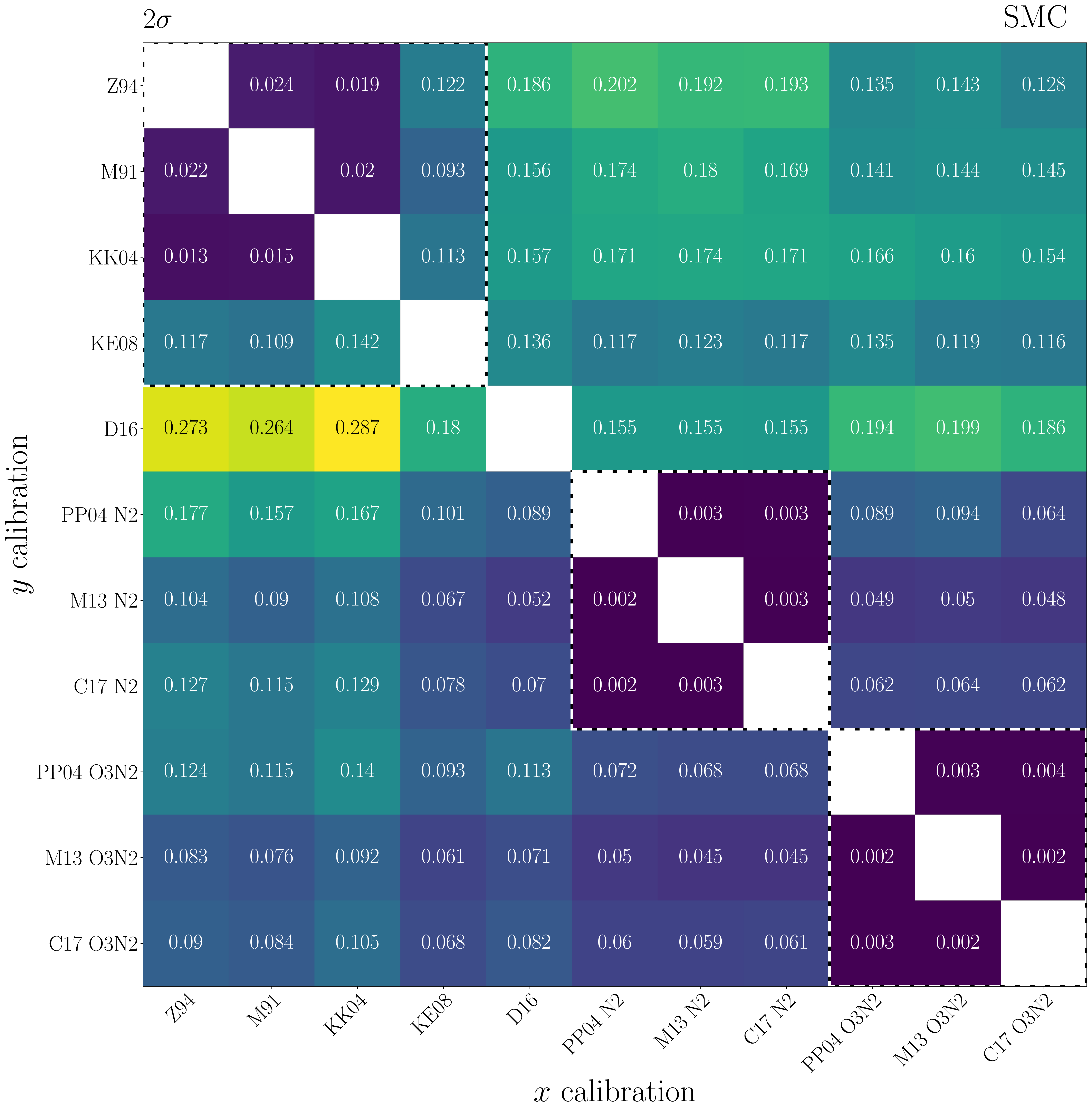}
  \caption{We present a summary chart indicating the 2$\sigma$ scatter surrounding our lines of best fit, for all calibration pairings ($x \rightarrow y$). The colorbar goes from dark purple for the lowest levels of scatter, to yellow for the highest values. Each square also has the 2$\sigma$ scatter, rounded to 3 decimal places, presented in the centres of each square. In black and white dashed squares, we show the regions which contain calibrations based on the same emission line ratios. The largest scatter is KK04$\rightarrow$ D16, at a 2$\sigma$ scatter of 0.287 dex. The smallest scatters are found between the N2 calibrations, and between the O3N2 calibrations.}
   \label{fig:scatter_matrix}
\end{figure*}

To summarize the scatter surrounding each of our polynomial fits, we show in Figure \ref{fig:scatter_matrix} all possible combinations of $x \rightarrow y$ calibration conversions. Each square is color-coded by the average of the positive 2$\sigma$ and the absolute value of the negative $2\sigma$ bounds, and this $2\sigma$ value is also shown at the centers of each square. Dark purple indicates very low scatter, and yellow, the largest scatter around our polynomial fits. In black and white dashed squares, we denote the regions where the calibrations are based upon the same emission lines. These boundaries, consistent with our analysis of Table \ref{tab:eline_sort}, contain the lowest scatter conversions between calibrations.

From this analysis, we conclude that the conversion between metallicity calibrations with very few lines in common is likely to be subject to higher scatter than conversions where a larger number of lines are in overlap. Converting between any two calibrations where the only required emission line in common is \NII, or conversions where H$\alpha$ is excluded from the $y$ calibration is particularly prone to larger scatter around a polynomial fit. However, differences in required emission lines are not the only controlling factor, as evidenced by large $2\sigma$ scatter being found between two R$_{23}$-based calibrations.

\subsubsection{Redshift dependence}

To further investigate potential sources of the scatter around our polynomial fits within the MaNGA sample, we can test the influence of covering fraction on the scatter present in these calibrations by dividing the MaNGA sample into two redshift bins at the median redshift, $z = 0.037$. This higher redshift and lower redshift subsample should be relatively free from physical evolution, and so the main difference between these two samples should be the covering fraction of the MaNGA fibres. The median redshift of the low redshift slice is $z=0.026$, which corresponds to 1.05 kpc coverage within the 2 arcsec fibre. The high redshift subsample corresponds to a typical physical scale of 2.2 kpc fibre$^{-1}$. 

We repeat the entire fitting process on both subsets of the data, and examine differences in the polynomial fits to the redshift selected subsample. To assess the difference between the two samples, we measure the median offset between the polynomials across the entire range of $x$ calibration. A large number of calibrations do not show substantial offsets between the higher and lower redshift bins; in particular we can expect those conversions with minimal scatter (e.g., C17 N2 $\rightarrow$ PP04 N2 or C17 N2 $\rightarrow$ M13 N2, as illustrated in Figure \ref{fig:scatter_matrix}) to not display redshift offsets. 

Typical 2$\sigma$ scatter for the MaNGA sample is 0.10 dex. We select all conversions which have a redshift offset in their polynomials of greater than $\pm 0.05$ dex, about half of the typical 2$\sigma$ scatter across all conversions. 
Only five out of our 110 conversions are prone to shifts $>0.05$ dex in their median offset when sliced by redshift. These conversions are: M91 $\rightarrow$ PP04 N2, C17 O3N2 $\rightarrow$ D16, Z94 $\rightarrow$ D16, KK04 $\rightarrow$D16, \& M91 $\rightarrow$ D16. These five notably also have larger than average $2\sigma$ scatter, so this represents a small relative shift for these conversions.

The redshift offsets are universally well below the 2$\sigma$ median scatter for all conversions.  Typical offsets between the polynomials for high and low redshift fits are 0.003 dex, more than an order of magnitude smaller than the typical scatter. Calculated on a conversion by conversion basis, the distinction between redshift polynomials represents only $\sim$3\% of the typical scatter.  Where larger offsets between redshift samples exist, they are usually strongest at the low metallicity end of the upper branch in R$_{23}$ conversions. 
Given the extremely small shifts overall when we divide our sample by redshift, we conclude that the redshift distribution, and thus the covering fraction, is not responsible for the varying level of scatter in these conversions.

Within MaNGA, we conclude that scatter around our polynomial fits to convert between any two metallicity metrics is generally minimized when the two calibrations have a large fraction of their emission lines in common. However, this overlap in emission line is not sufficient to guarantee minimal scatter, as not all of the R$_{23}$-based calibrations convert smoothly into each other. These varying levels of scatter, even with consistent emission lines, are likely due to differences in the underlying theoretical models. Redshift (covering fraction) is not a driving factor for scatter between any two calibrations in the MaNGA sample. 

\subsection{Consistency with other data sets}

While the polynomial fits presented here are valid for the current DR15 MaNGA sample, we assess whether they can be flexibly applied to other data sets beyond the MaNGA data used here. 
We make use of two comparison data sets, which cover a wide range of spatial resolution, to assess the viability of using these conversions on other data sets. 

\subsubsection{SDSS I/II comparison sample: 3 kpc spatial scale} 
First, we use the work of \citet{Teimoorinia2021}, which reassessed the work of \citet{Kewley2008} using the larger data volume available in the SDSS DR7. In this SDSS DR7-based work, S/N $> 8$ is required in the emission lines for each metallicity calibration, and a redshift limit $0.04 < z < 0.1$ is imposed. T21 use an SMC-like dust correction curve \citep{Pei1992}, which is assessed to have negligible impact on the resultant metallicities compared to a MW-like dust correction curve, and is consistent with the SMC values used in the current work. Roughly 61,000 galaxies had calculable metallicities; each of these galaxies had a single, 3 arcsecond spectrum taken at the nucleus of each galaxy. T21, along with providing publicly available Random Forest models, produced 3rd order polynomials for conversions between an identical set of metallicity calibrations used in the present work. 

\subsubsection{TYPHOON comparison sample: 100 pc spatial scale}
Second, we use a set of 4 galaxies observed as part of the TYPHOON IFS survey (Seibert, Rich \& Madore, 2021, private communication): NGC 2997, NGC 5068, NGC 5236 (M83), \& NGC 2835. These galaxies range from a redshift of $z=0.00171$ to $z=0.003633$, placing them at much lower redshifts than the MaNGA sample, which has a median redshift of $z \simeq 0.03$. TYPHOON is observed with a 1.65$^{\prime\prime}$ slit, which corresponds to physical scales ranging from 57 pc for NGC 5236\footnote{Differences in scale between this estimate and the 45 pc quoted in \citet{Poetrodjojo2019} are due to the change from Galactocentric Hubble flow distances to cosmological distances.} to 123.75 pc per spaxel for NGC 2997. This is an order of magnitude smaller in resolution element than the typical 1.8 kpc scales of the MaNGA survey \citep{Law2016}, and a factor of $>$25 smaller than that of the SDSS DR7. For the current comparisons, we use the data cubes at their native resolutions, and we summarize the data (redshifts, spatial scale in parsecs, and the number of spaxels for which any metallicity is calculable) in Table \ref{tab:typhoon}.

\begin{table}
	\caption{Summary of the TYPHOON data used in this work. We show the galaxy NGC number, the redshift, the corresponding spatial scale of the spectra, and the number of spaxels which have at least one calculable metallicity.}
	\label{tab:typhoon}
\begin{tabular}{l|ccc}
\hline
Galaxy  & redshift ($z$) & resolution (pc) & $n$ spaxels\\
  \hline
  NGC 5236 &0.00171& 57   & 3,681 \\
   NGC 5068 & 0.002235& 76  & 2,583 \\
    NGC 2835 & 0.002955&101  & 1,529 \\
 NGC 2997  &0.003633 &124 & 1,526 \\
 \hline
\end{tabular}
\end{table}

To calculate metallicities for the TYPHOON galaxies, we begin with data cubes of emission line strengths, and manipulate these emission line fluxes in an identical fashion to the MaNGA data. We impose the same data requirements: S/N $> $5 in all required lines, flagged as SF on the BPT diagram via \citet{Kauffmann2003} or \citet{Stasinska2006}, and dust corrected via the SMC dust correction curve. This data processing is slightly more stringent than that of \citet{Poetrodjojo2019}, which required S/N > 3 in all lines\footnote{\citealt{Poetrodjojo2019} also used the MW dust correction curve. However, this choice of MW or SMC curve, based on comparisons between DR4 and DR7 \citep{Teimoorinia2021}, should not influence these results.}. Our data processing generates a set of 11 metallicity values which are based entirely on the four TYPHOON galaxies, and conversions between these metallicities can be undertaken in a consistent manner. We use the same fitting procedure to compare metallicities in TYPHOON spaxels, with only two changes. 

The first is a reduction in the number of horizontal bins from 100 to 50 in the determination of the percentile ranges, which reflects the lower data volume in the TYPHOON data set of 9,319 spaxels with at least one calculable metallicity. The number of viable spaxels per galaxy is presented in Table \ref{tab:typhoon}. Secondly, we then fit 3rd order polynomials instead of 5th order to the TYPHOON data, for the same reason of reduced statistical power. The TYPHOON data has sufficiently poor number statistics in the lower branch of R$_{23}$-based calibrations (typically fewer than 20 spaxels) that the lower branch is excluded from the analysis which follows. For calibrations involving the R$_{23}$ line ratio, due to the paucity of \OII~detections (also noted in \citealt{Poetrodjojo2019}), the number of viable spaxels is severely limited and typically numbers $\sim 600$ across all four galaxies. Calibrations which do not involve this line ratio typically have $>8000$ spaxels with calculable metallicities.

\subsubsection{Comparisons between polynomials}
\label{sec:poly_datasets}
\begin{figure*}
  \includegraphics[width='7in]{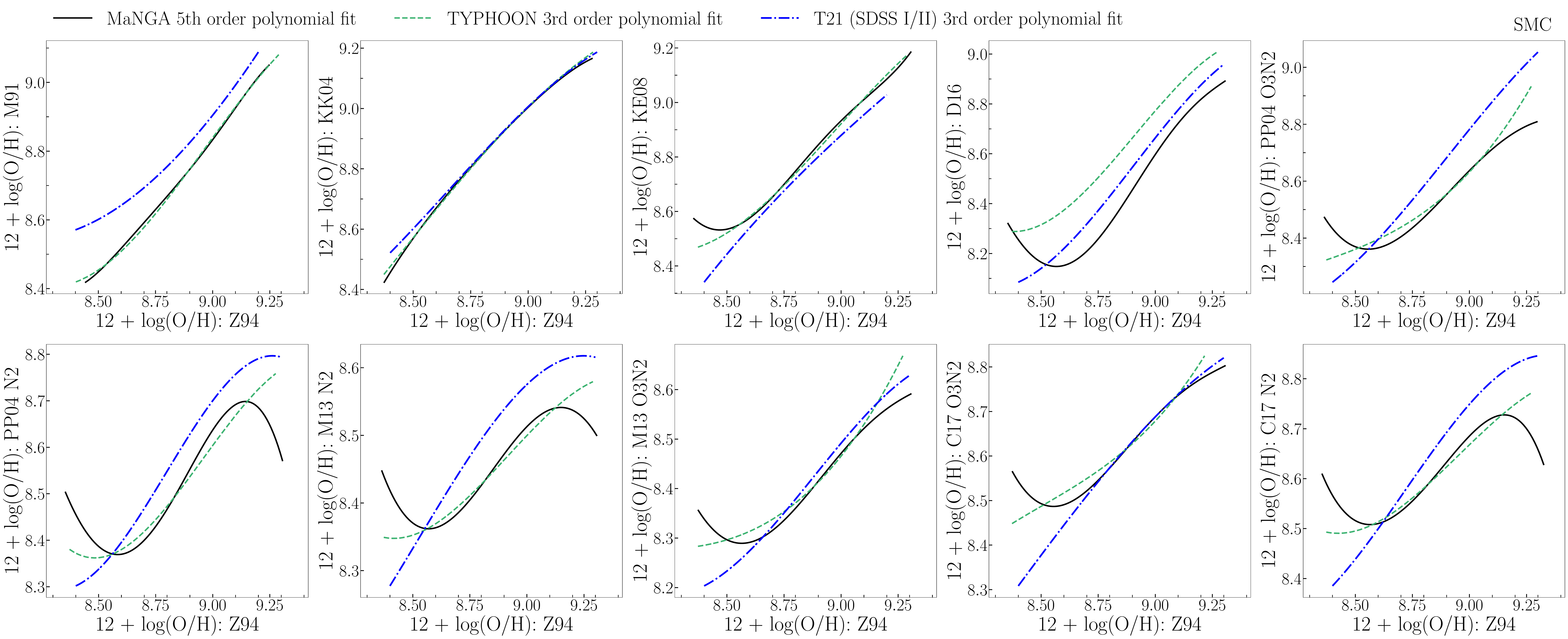}
  \caption{Comparison of the polynomial lines of best fit. The 5th order polynomial fit to the MaNGA data presented here are plotted in a black solid line. We plot the 3rd order T21 polynomials in a blue dot-dashed line. 3rd order polynomial fits to the TYPHOON data are plotted in a dashed green line.}
  \label{fig:z94_poly}
\end{figure*}

Given that these three data sets probe very different spatial scales, comparing between them will allow us to assess scale-invariance of these  conversions, between the $\sim$100 pc regime of TYPHOON, the $\sim$1.8 kpc resolution of the MaNGA data, and the $\sim$3 kpc nuclear spectra of the SDSS DR7. We plot the polynomials which convert between metallicities on the same axes\footnote{In some of the supplementary Figures, there are several calibrations which have discrepancies in the metallicity at which they transfer from upper to lower branches. In particular, there are several KK04 calibrations which in the current work have a strong transition from upper and lower branch at 12+log(O/H) = 8.4, but which are fit across this boundary in \citet{Teimoorinia2021}. In the DR7, the upper and lower branch are naturally fit by a third order polynomial for certain conversions, and so no separate lower branch was implemented. In the MaNGA data, this division is significant, and imposed for all conversions.} in Figure \ref{fig:z94_poly}.

In Figure \ref{fig:z94_poly}, we plot the MaNGA based curve in a black solid line, the DR7-based polynomials in a blue dot-dashed line, and the polynomial based on the TYPHOON data in a green dashed line, for the conversion from Z94 into all other calibrations. 
We note immediately that a subset of the conversions appear to have excellent agreement between all three data sets; in particular, Z94 $\rightarrow$ KK04 \& Z94 $\rightarrow$ KE08. Z94 $\rightarrow$ M91 has excellent agreement between MaNGA and TYPHOON. 

\begin{figure*}
\includegraphics[width=.495\textwidth]{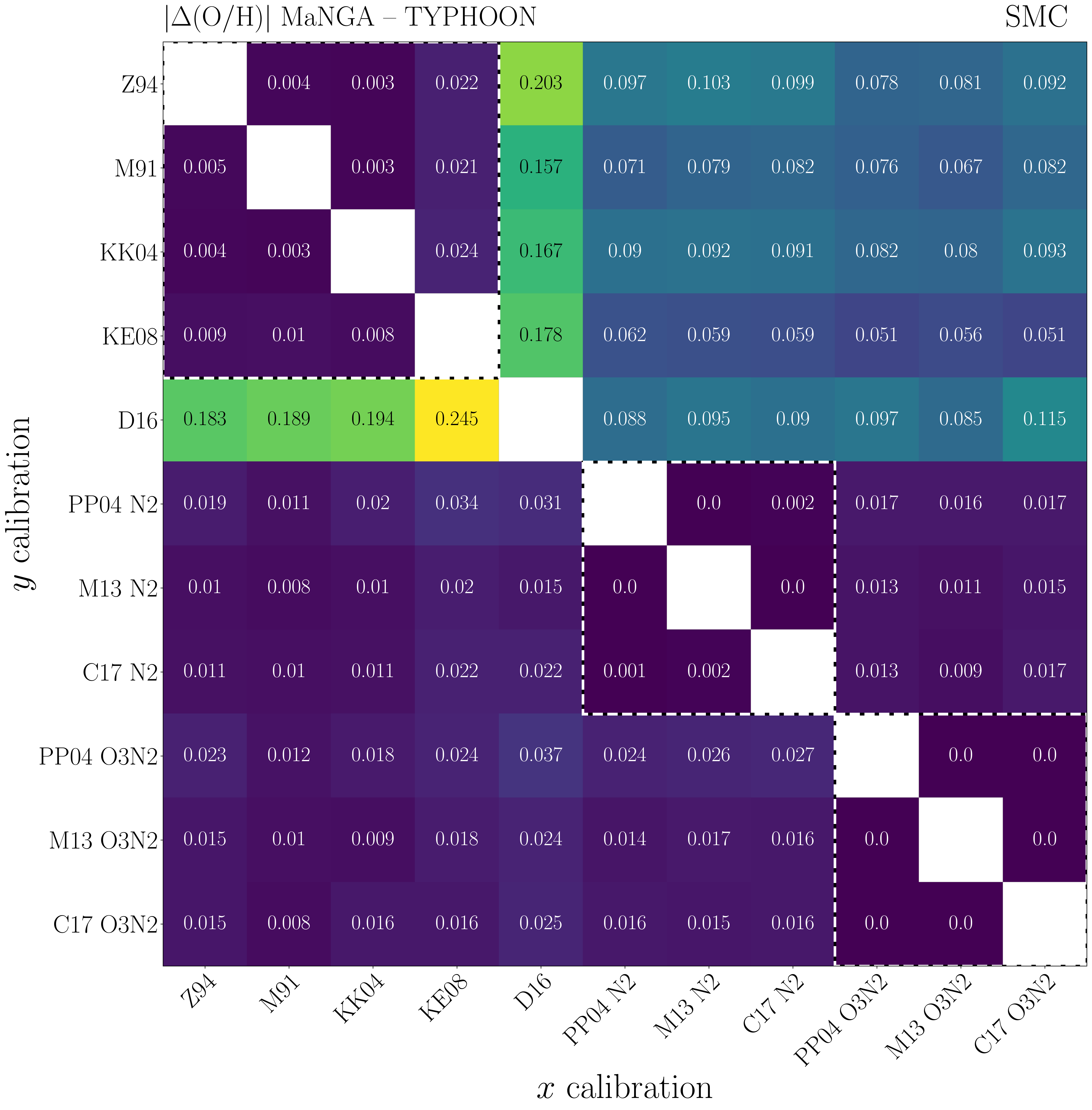}
  \includegraphics[width=.495\textwidth]{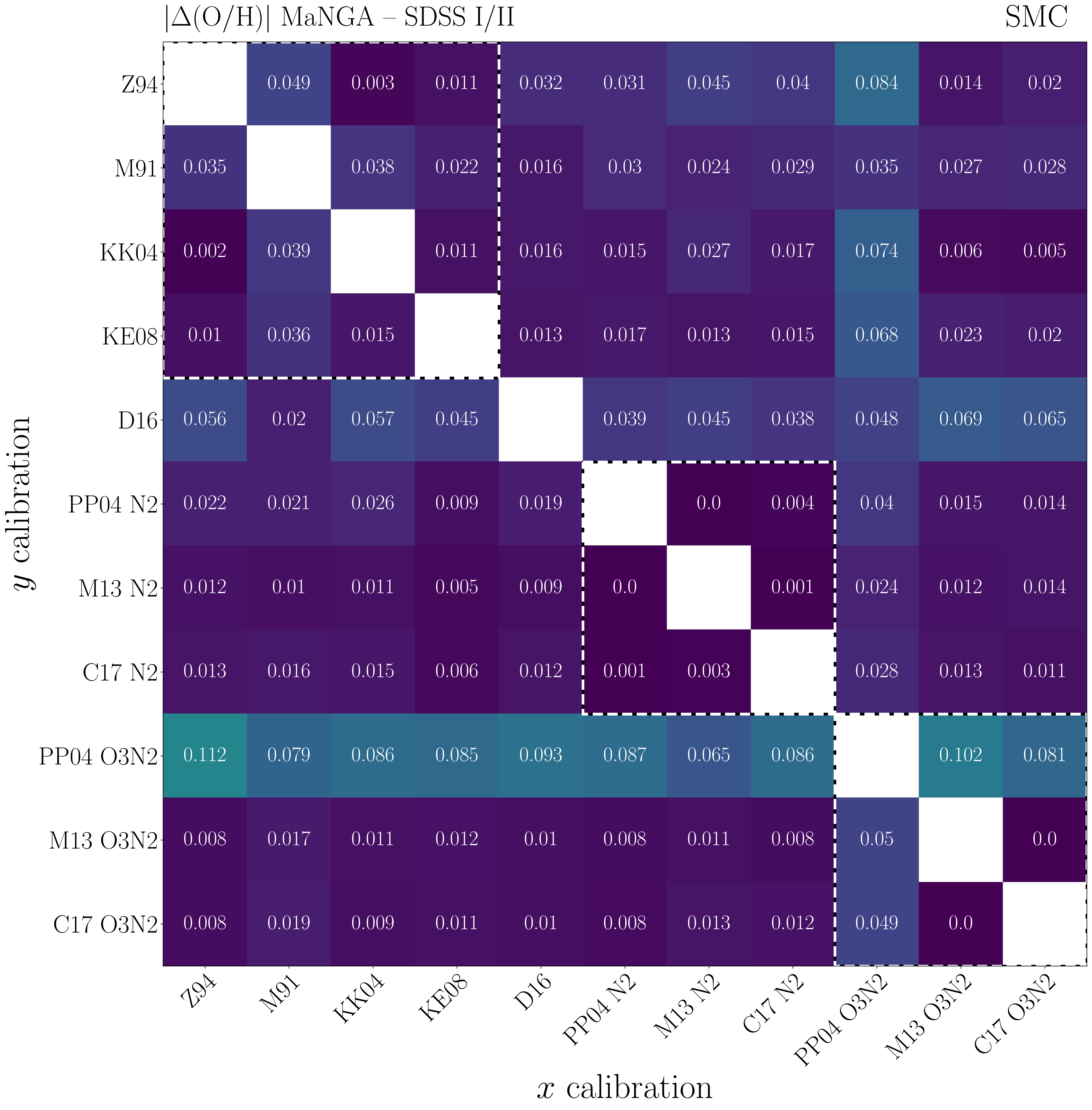}
  \caption{Comparison of the differences between polynomial lines of best fit. In the left hand panel, we take the absolute value of the difference between the MaNGA polynomial and the TYPHOON polynomial, and in the right hand panel, the absolute value of the difference between the MaNGA polynomial and the DR7 polynomial. These differences are calculated across the range where both polynomials are valid. In each square we indicate the median offset between the polynomials across both data sets; both panels have the same colorbar to indicate the magnitude of the differences between polynomials. To show an offset of 0.0, the polynomials must be in perfect agreement. The median difference between the MaNGA \& TYPHOON polynomials (left) is 0.019 dex, while the median difference between MaNGA \& DR7 (right) is 0.017 dex.}
\label{fig:superfig_polyoffset}
\end{figure*}

We present in Figure \ref{fig:superfig_polyoffset} an overall summary of the offsets between these polynomial fits, similar to Figure \ref{fig:scatter_matrix}. For each calibration pairing, we identify the $x$ range where all three polynomials are valid. We then sample this $x$ range evenly, and produce $y$ values from each polynomial. We then subtract the $y$ values from the TYPHOON polynomial from the MaNGA polynomial, and separately subtract the DR7 polynomial results from the MaNGA polynomial. We then take the absolute value of all differences, and find the median offset for each pair of polynomials (MaNGA - TYPHOON \& MaNGA - DR7), in the left and right panels of Figure \ref{fig:superfig_polyoffset} respectively. 
We show the results of this test for all calibration conversions in Figure \ref{fig:superfig_polyoffset}, where the colorbar is scaled to the median polynomial offset, and each offset value is indicated in the centre of each cell. The colorbar is scaled identically in both the left and right hand panels. For the MaNGA-TYPHOON data pairing, it is clear that offsets between the TYPHOON and MaNGA data are much higher for conversions between D16 and any of the R$_{23}$-based calibrations than for any other pairing. The largest discrepancies between any of the three polynomials are $\sim 0.25$ dex, for KE08 $\rightarrow$ D16, and $\sim$ 0.18 dex for KK04 $\rightarrow$ D16, and Z94 $\rightarrow$ D16, and all between the MaNGA and the TYPHOON data. The differences between MaNGA and the DR7 are comparable in magnitude, with the exception of having smaller polynomial offsets when converting between D16 and any of the R$_{23}$-based calibrations. The typical offset between MaNGA and TYPHOON polynomials is 0.019 dex, and the typical offset between MaNGA and DR7 is 0.017 dex. In both cases, a substantial number of the polynomials have offsets less than 0.01 dex. 

We note that while MaNGA - TYPHOON does show higher offsets between polynomial fits for certain calibration conversions than the MaNGA - DR7, in other cases, MaNGA - TYPHOON shows smaller offsets. The conversions between R$_{23}$ based calibrations, and conversions from \& to PP04 O3N2, are generally more consistent between MaNGA and TYPHOON than they are between the MaNGA and DR7. This is interesting both because the difference in spatial scale is larger between MaNGA and TYPHOON than it is between MaNGA and the DR7, and because the survey design is more similar between MaNGA and DR7. However, in contrast, the MaNGA - TYPHOON polynomial offsets are higher for N2-based and O3N2-based calibrations when converting into R$_{23}$ based calibrations than they are for MaNGA - DR7, so we have not uncovered a universal trend. 

In comparing Figures \ref{fig:scatter_matrix} \& \ref{fig:superfig_polyoffset}, we note some similarities between which conversions have minimal scatter within the MaNGA sample exclusively, and the differences between the polynomials fit to entirely different data sets. For the MaNGA sample, we noted that there is functionally zero scatter for around conversions between calibrations when both $x$ \& $y$ calibration are based either on the O3N2 line ratio or on the N2 line ratio. Polynomials fit to entirely different data sets, for these same set of conversions (among N2-based calibrations, or among O3N2 calibrations) are functionally identical. 

The similarity between Figures \ref{fig:scatter_matrix} \& \ref{fig:superfig_polyoffset} also indicates that the conversions with lower $2\sigma$ scatter, discussed in Section \ref{sec:DR15_elines} to be minimized when the emission line ratios rely on a consistent set, are also the conversions where the conversions are most consistent across data sets. In T21, systematic changes in the emission line fluxes between the DR7 and DR4 were found to influence the positioning of the polynomial conversion between the two data sets. While most of the emission lines were revised upwards, the per cent change was not consistent across all emission lines. As a result, calibrations which relied on consistent emission lines were more consistent between the data sets, than those which were subject to this additional change. We thus caution that when using these polynomial functional forms, the details of the data processing may cause systematic changes. 

It is also worth noting that the magnitude of the offsets between polynomials fit to different data sets is typically substantially smaller ($\sim 3\times$ for MaNGA - TYPHOON \& $\sim 5\times$ for MaNGA - DR7) than the $2\sigma$ scatter around the MaNGA data alone. Even for one of the largest polynomial offsets between data sets (Z94 $\rightarrow$ D16, 0.183 dex), the $2\sigma$ scatter for that particular conversion in MaNGA alone is 0.273 dex, 1.5$\times$ as large. Nearly every entry in Figure \ref{fig:scatter_matrix}, showing the $2\sigma$ scatter, is larger than the corresponding entry in Figure \ref{fig:superfig_polyoffset}, showing the differences between polynomials fit to different data sets, indicating that these polynomial offsets between data sets are much smaller than the scatter around a given conversion. In the supplementary material, we include the residuals of the polynomial fits shown in Figure \ref{fig:z94_poly}, subtracted from the MaNGA polynomial. We denote the MaNGA $2\sigma$ range in a pale blue background shading. 

In 16 cases across both data set comparisons, there are larger offsets between data sets than $2\sigma$ scatter around the MaNGA curves. 11 of these are found in MaNGA-DR7; and almost all are cases where the scatter around the MaNGA polynomials is near zero, and the offsets between polynomials fit to the different data sets remains low ($\sim 0.02$ dex) even in these cases. The only major exception is M13 O3N2 $\rightarrow$ PP04 O3N2, where the scatter is near zero in MaNGA, but the DR7 polynomial is offset by 0.1 dex.  
The remaining 5 are for MaNGA-TYPHOON; these are D16 $\rightarrow$ R$_{23}$ calibrations and KE08 $\rightarrow$ D16. We return to a potential explanation for these high offsets in Section \ref{sec:discussion}. 

We conclude from these initial tests that the polynomial fits presented here, based on MaNGA, are suitable for use in other data sets. In general, the differences in polynomial fits between those fit to the MaNGA data, and those fit to the DR7 or to TYPHOON, are minimal (0.019 dex). These polynomial fits are therefore functional over the redshift range probed here ($0.00171 < z < 0.1$), which corresponds to a spatial resolution range of 57 pc - 5.5 kpc. 

\subsubsection{Spatial resolution blurring}

In spite of the remarkably good agreement between polynomials fit to very different data sets, we wish to touch on two issues of particular concern when comparing data of very different spatial resolution. The first is the impact of simple spatial blurring on the metallicity measurements.

Low spatial resolution blurs together many separate \HII~regions, leading to an averaged metallicity measurement; and certainly this blurring will influence the \textit{absolute} measurements of a metallicity. As we proceed from low resolution (3 arcsecond fibre, 0.04 < $z$ < 0.1 ) to high resolution (1.65 arcsec, 0.00171 < $z$ < 0.0036), we can expect that the raw metallicity values calculated may vary. If this were a dominant effect in the conversions between calibrations, we would expect to see that the MaNGA, DR7, and TYPHOON polynomials would not align well with each other, that this discrepancy would be strongest between the DR7 and TYPHOON data, and that offsets to our calculated conversions would affect all calibration pairs. 

However, we see from Figure \ref{fig:superfig_polyoffset} that in most cases, the DR7, MaNGA, and TYPHOON curves are in excellent agreement despite a varying spatial resolution of nearly two orders of magnitude. The DR7 data corresponds to a typical fibre coverage of $\sim 3$ kpc, while the TYPHOON data data corresponds to a physical scale of $\sim100$ pc. We conclude by the agreement between some calibrations, though it is not universal, that pure spatial blurring does not dominate the conversions between calibrations. This lack of sensitivity to pure spatial resolution indicates that either all metallicity calibrations are roughly equally affected by the resolution effects, and/or that the gas-phase metallicity is not substantially varying on these spatial scales. A few studies of 2 dimensional maps of the variation in metallicities across the disk of a galaxy have appeared in the literature; \citet{Kreckel2020} finds increased homogeneity in the gas-phase metallicity at sub-kpc scales in 8 galaxies in the PHANGS-MUSE survey, while \citet{Li2021} finds a mean correlation length of 1 kpc in 100 galaxies in the CALIFA IFS survey.  While this spatial scale is smaller than the roughly 3 kpc scales of the DR7, it would help to tie together the 1.7 kpc scales of MaNGA and the 100 pc scales of TYPHOON. 

\subsubsection{The influence of diffuse ionized gas}

Our second issue in comparing between data sets is the potential presence of diffuse ionized gas (DIG). The role of DIG light in metallicity calculations has become a subject of considerable concern, as it is preferentially a low surface brightness feature \citep[e.g.,][]{Reynolds1984, Reynolds1990, Haffner2009}, but due to its ubiquitous nature, can be increasingly flux-dominant as the spatial scale of the spaxel increases \citep{Poetrodjojo2019}. The DIG light tends to also become more dominant at large radii \citep{Poetrodjojo2019}, so centrally situated fibres such as those which are present in the DR7 will be less impacted by DIG light than the larger radius spaxels present in MaNGA. As the DIG light is not produced by photoionization \citep{Zhang2016}, metallicity calibrations which rely on photoionization models will not be well adapted to this diffuse light. The DIG's emission pathway thus results in alterations to the emission line ratios used to calibrate metallicities.

How strong an impact this ought to have on the final metallicities seems to depend on both how DIG-dominated spaxels are identified, and what the quality control process involves. \citet{Pilyugin2018} assessed the validity of the metallicity in MaNGA spaxels, given that there might be significant DIG-based biases in the data, finding broadly that the inclusion of BPT-based cuts will remove the majority of the most severely affected spaxels, as it is in the `composite' region of the BPT diagram where the DIG-dominated spaxels are most likely to reside. \citet{Law2020} find that for the BPT diagram used in the present work (\OIII/H$\beta$~vs \NII/H$\alpha$), the \citet{Kauffmann2003a} curve, in combination with a S/N $> 5$ cut, does a good job of separating dynamically cold regions, such as those found in \HII~regions in the Milky Way, from dynamically warm regions. \citet{Zhang2016} has suggested that if evolved stars are responsible for some of the production of DIG light, the spectrum would shift into this composite region of the BPT diagram, reflecting a harder ionization field. In addition, \citet{Law2020} find that a selection criterion limiting H$\alpha$ EW $< 2.0$ excludes the dynamically cold region. Therefore, our selection criteria of H$\alpha$ EW $> 6.0$ \AA~should also be excluding a large component of dynamically warm, LINER-like regions, even if they had passed the BPT criterion.

Although our selection criteria are likely to exclude the majority of DIG-dominated spaxels, there may still be some remaining contamination, so it is wise to assess the magnitude and direction of low-level contamination on our metallicity conversions.
\cite{Zhang2016} came to the conclusion that DIG light will typically overestimate metallicity in MaNGA. The DIG component has a lower ionization parameter than an \HII~region associated with star formation, and as metallicity calibrations rely on a relationship between the ionization parameter and the metallicity, in DIG-dominated regions, this correlation breaks down and the calibration will fail. The impact of this bias changes depending on the line ratio used for the metallicity calibration.
O3N2 calibrations and D16 are found in \citet{Zhang2016} to vary by $\pm$ 0.05 dex. R$_{23}$-based calibrations typically are biased low by $-0.1$ dex, and N2 by 0.05 - 0.1 dex high. Any DIG-dominated spaxel should be DIG-dominated for all calibrations; it is possible for R$_{23}$-based calibrations and N2-based calibrations to have their biases cancel each other, but for all other combinations, we should expect to see increased scatter in the conversion, as a result of any remaining DIG contamination. 

By way of example; any of our calibrations will capture some set of spaxels which are truly dominated by \HII~regions. These spaxels will produce some relation between a chosen $x$ calibration and any other. For the sake of simplicity, let us assume the  $x$ \& $y$ calibrations are both N2-based. DIG dominated spaxels will then be offset from these \HII~region-dominated spaxels, biased high by 0.05 - 0.1 dex. For the $x$ calibration, this would extend the DIG-dominated spaxels to higher $x$ values. The same would be true of the $y$ metallicity. Since the bias in this case is the same for both target and source, we would expect that the conversion between them would be shifted above the conversion line for \HII~dominated spaxels, diagonally up and to the right. In the case where the calibrations follow a 1-1 line, then this bias could be invisible, as the DIG-dominated spaxels would simply move up along the line. However, the N2 $\rightarrow$ N2 conversions we present here are not 1-1 linear functions, and so we would expect to see that the presence of DIG-influenced spaxels would result in an increase in scatter. 

In contrast to this expectation, N2 $\rightarrow$ N2 calibrations, along with O3N2 $\rightarrow$ O3N2 calibrations, present almost zero scatter, in spite of their nonlinear relationships. We can conclude from this non-detection of substantial scatter that we are unlikely to be working with a large volume of DIG-dominated spaxels, in line with our expectations as a result of our S/N cuts and BPT classifications. 

To check directly for the influence of DIG-dominated spaxels in MaNGA, we select, for all metallicity pairings, the number of spaxels which would be classified as DIG-dominated by \citet{Poetrodjojo2019} using log(\SII/H$\alpha$) > 0.29. 
We find that this number ranges from 7 spaxels to 98 spaxels, with a typical selection of 97 spaxels, representing, typically, 0.25 per cent of the overall sample. There is a 0.2 per cent dropout rate due to non-detections in the \SII~line. \HII~regions therefore typically comprise 99.75 per cent of the sample in the MaNGA data. If we repeat this process for the TYPHOON sample, we find that zero spaxels with metallicities pass the DIG-dominated classification of \citet{Poetrodjojo2019}, with a median dropout rate of 0 per cent where \SII~is nondetected.  In the DR7, there is again a 0.2 per cent dropout rate for \SII~non-detections, and the HII-dominated fraction is 99.99 per cent, with somewhere between zero and 8 spectra flagged as DIG-dominated. 

We thus do not expect to be influenced by a substantial fraction of DIG-dominated spaxels, given our BPT and S/N requirements for each spaxel. This negligible DIG-dominated contamination is consistent with the results of both \citet{Law2020} and \citet{Zhang2016}, where a combination of BPT classification and high S/N cuts remove a majority of DIG-dominated spaxels.  \citet{Poetrodjojo2019} did not perform a BPT classification prior to assessing metallicities; so while these metallicities can be calculated, they are likely to be affected by multiple biases, one such being the inclusion of more DIG spaxels. We find that our sample is $> 99$ per cent HII-dominated regions, and do not expect low-level contamination to influence the metallicity calibrations presented here.

\section{Discussion}
\label{sec:discussion}

We have provided 5th order polynomial fits to faciliate conversions between a set of 11 different gas-phase metallicities, based upon roughly 1.1 million spaxels for each metallicity calibration. We assess relative levels of scatter within our MaNGA DR15 sample, and find that the scatter around the polynomial fit is minimized when the $x$ \& $y$ calibrations are based on similar sets of emission lines. We determine that redshift does not influence the positioning of these fits within MaNGA. We additionally compare the polynomial fits to two different data sets, the SDSS DR7, and the IFS survey TYPHOON, which provide a total dynamic range of nearly two orders of magnitude in resolution, from $\sim 50$ pc to 5.5 kpc. We find that the polynomial fits are in excellent agreement across data sets, and we do not expect any bias from the inclusion of DIG-dominated spaxels in any of the data sets, nor do we see any evidence for strong effects based purely on spatial resolution blurring.  

\subsection{Galactocentric radius effects}
\begin{figure*}
  \includegraphics[width=\textwidth]{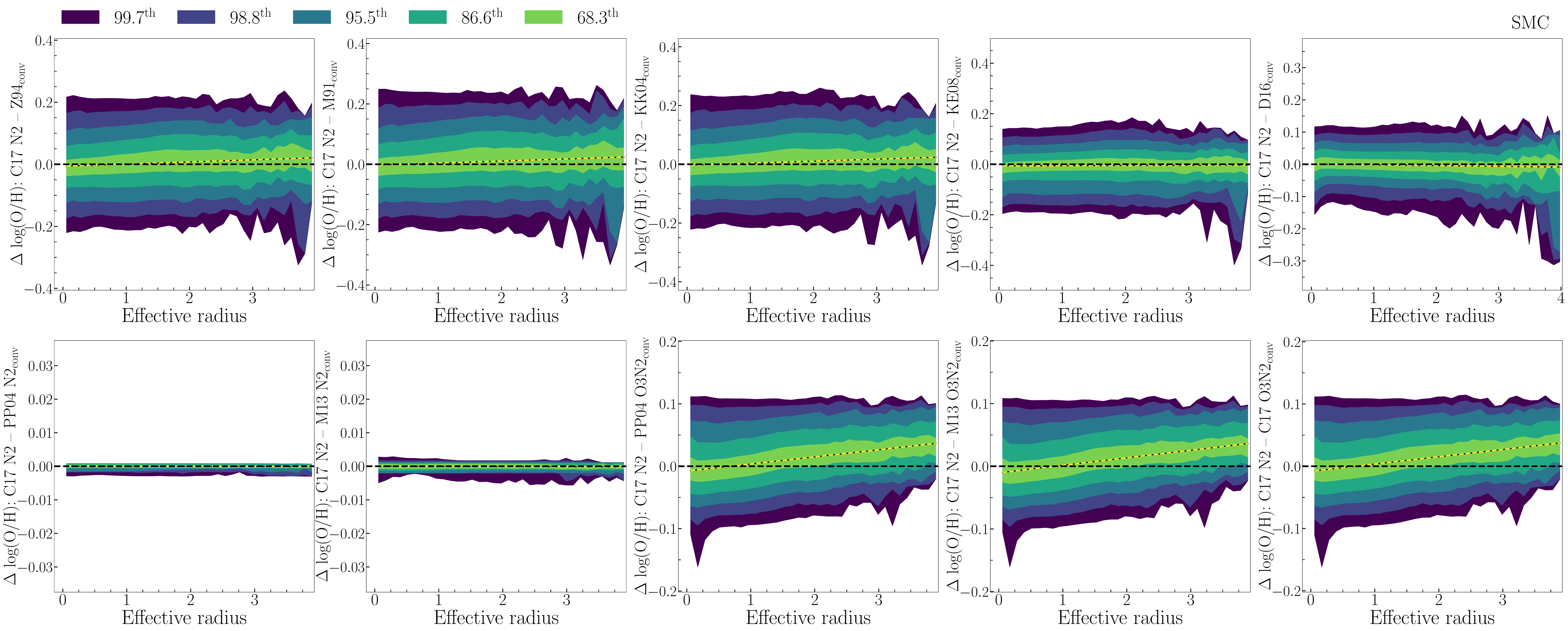}
  \caption{The results of plotting the effective radius for all spaxels in MaNGA ($x$) vs the offset from the polynomial best fit ($y$) for all calibrations converting into C17 N2. Colored regions indicate, as with previous similar Figures, the percentile bounds for 100 horizontal bins. The data is truncated at 4R${_e}$. A horizontal black line indicates a line of zero slope, and a yellow-black dashed line indicates the line of best fit to the 1$\sigma$ contours in light green. These data do not show complex morphologies, so a first order polynomial fit is sufficient.}
   \label{fig:re_ex}
\end{figure*}

As part of the change from single-fibre, nuclear metallicities, to those provided by IFS surveys, we now have many metallicity values for each galaxy. 
We can, for the first time, investigate the role that galactocentric radius may have on the conversions between calibrations in MaNGA. For all calibration pairs, we find the offsets of all spaxels in the MaNGA sample from the polynomial fit, and the effective radius of the spaxel, corrected for inclination angle. We then perform the same percentile bounds finding exercise as for the calibrations (e.g., Figure \ref{fig:intoC17N2}) for the effective radius $R_e$ vs the vertical offset from the polynomial fit. We show an example of the results of this test in Figure \ref{fig:re_ex}, for all calibrations converting into C17 N2. 

\begin{figure} 
\includegraphics[width=\columnwidth]{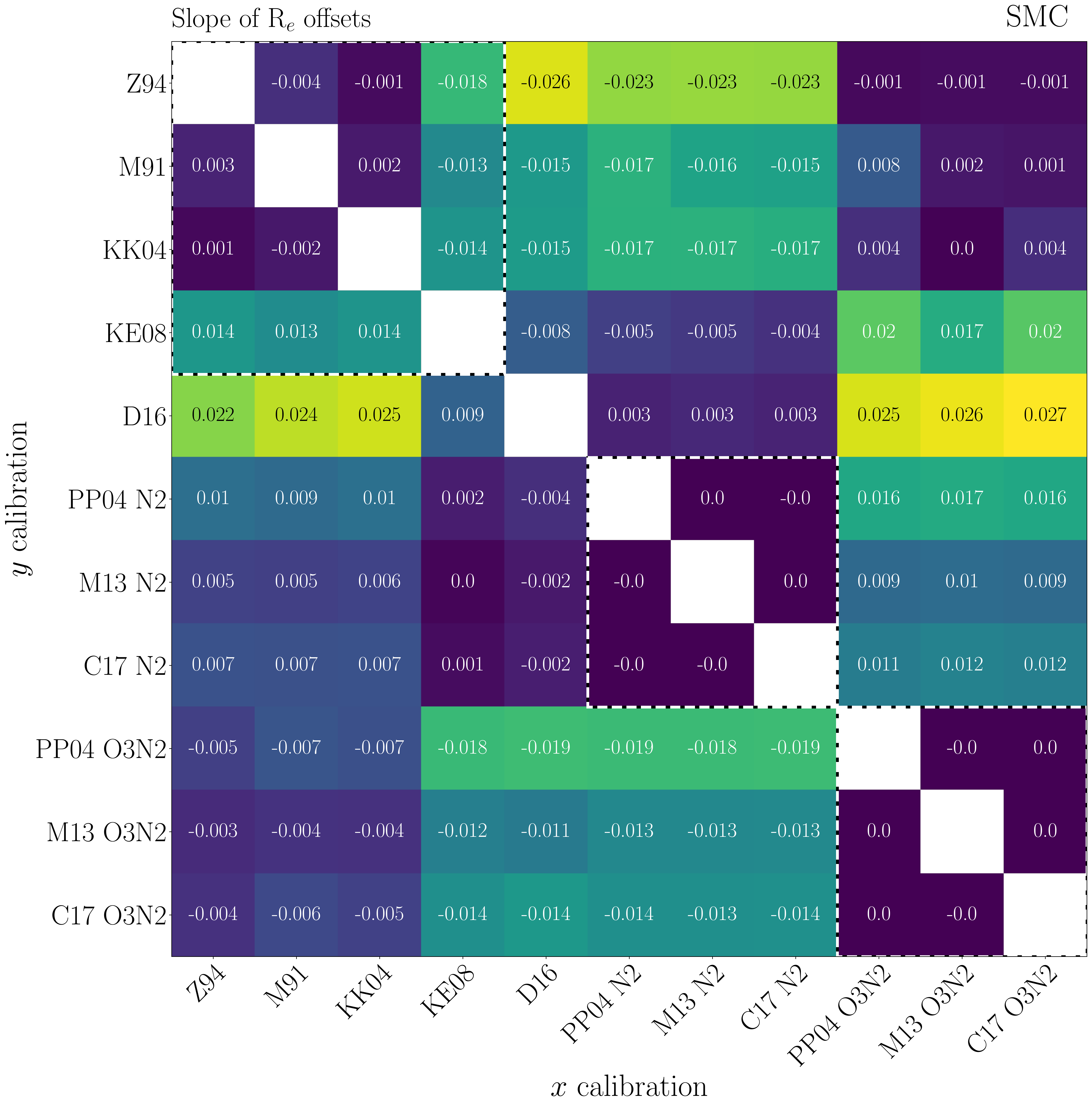}
  \caption{For galaxies in MaNGA, we plot the slope of a linear fit to the relationship between the offset from the polynomial fit and the effective radius R$_e$, corrected for inclination angle. The typical magnitude of the slope is 0.008 dex R$_e$$^{-1}$, very nearly zero, though the maximum slope has magnitude 0.027 dex~R$_e$$^{-1}$. }
  \label{fig:DR15_Re_slope}
\end{figure}

We limit effective radius to less than 4$R_e$, as number statistics become extremely poor outside that region. We fit a first order (linear) fit to the 1$\sigma$ contours of the data, and from this acquire a slope of the relationship between effective radius and the offset from the polynomial fit. As seen in Figure \ref{fig:re_ex}, and confirmed with a visual inspection of all other calibration conversions, a first order line of best fit is sufficient. If there is no dependence on galactocentric radius, then we should fit a slope of approximately zero, i.e., that all offsets are equally likely to be found at all galactic radii. If certain calibration conversions, on the other hand, do show a trend with effective radius, we can capture the strength of this correlation by the slope of the line. These slopes are summarized in Figure \ref{fig:DR15_Re_slope}. As with similar figures, the color coding indicates the magnitude of the slope of the line, and the slope itself is indicated in the centre of each square. Dark purple indicates slopes close to zero, while yellow indicates higher slopes. 

Conversions between calibrations from and to consistent emission lines, which have very small $2\sigma$ scatter, also are independent of galactocentric radius, showing no relationship between metallicity offset from the polynomial, and effective radius. Across all calibration pairs, the median slope is 0.0 dex R$_{e}$$^{-1}$; the median magnitude of the slopes across all conversions is 0.008 dex R$_{e}$$^{-1}$, with a maximum of 0.027 dex R$_{e}$$^{-1}$.  Given the low typical dependence on R$_e$, we would not expect this to impact the majority of the conversions. However, for those conversions with slightly larger dependencies, (e.g., C17 O3N2 $\rightarrow$ D16), we find that of the 1,012,224 spaxels with R$_e$ < 4.0,  707,094 ($\sim$ 70 per cent) of them are found at < 1 R$_e$, and so should be subject to an offset bias maximally of order the slope (0.027 dex)\footnote{Repeating this slope-finding analysis including only spaxels which are found at < 1.5 R$_e$ does not substantially change these results. The median slope remains at 0.0 dex R$_{e}$$^{-1}$. The median magnitude of the slope increases very slightly to 0.014 dex R$_{e}$$^{-1}$.}. Given that the 2$\sigma$ scatter for C17 O3N2 $\rightarrow$ D16 is 0.186 dex, this slope maximally represents 14.5\% of the total scatter for spaxels far from the intercept.

We test the strength of this dependence directly. We take our polynomial fits, and for each spaxel, we correct for its R$_e$ by using the slope of the line fit to that conversion. We then find the $2\sigma$ scatter for these conversions, having been calibrated for their $R_e$ location. We can then compare the 2$\sigma$ scatter for the full MaNGA sample (as illustrated in Figure \ref{fig:scatter_matrix}) to the scatter in this newly corrected conversion. If the R$_e$ correction has no substantial impact on the scatter around a calibration, we should expect to see that the two $2\sigma$ values are approximately equal. If the R$_e$ correction is a substantial influence, we should see reductions in the scatter in the R$_e$ corrected sample. Subtracting the scatter of the R$_e$ corrected correlations from the full $2\sigma$ scatter results in a median change of 0.0006 dex\footnote{The median change when calibrated only to spaxels within R$_{e} < 1.5$ is 0.001 dex. The maximum reduction in scatter is unchanged: 0.02 dex. The choice of R${_e}$ limit therefore does not affect these results.}, with a maximum change of 0.02 dex. We therefore conclude that the effective radius of a given spaxel does not drive the scatter around the calibration conversions. 

\subsection{Galaxy-by-galaxy variability}
 In addition to the effective radius of a given spaxel, the IFS data gives us enough data on a per-galaxy basis that we can test if there is significant galaxy-to-galaxy variability in the conversions between calibrations. This particular test can only practically be done with MaNGA; with the sample of four galaxies from TYPHOON, we do not have any statistical power, but MaNGA, with 4824 unique galaxies, can be used to probe if some individual galaxies tend to lie above or below the polynomial fit describing the typical conversion between two metallicity calibrations.
 
 To assess whether individual galaxies are variable for our conversions between calibrations, we conduct a test similar to that illustrated in Figure \ref{fig:intoC17N2_resid}, which showed the offset of the contours of our data from the polynomial line of best fit. For each galaxy in the MaNGA sample, we select all spaxels, converting the $x$ metallicity into the $y$ metallicity. We then subtract the converted $x$ metallicity from the $y$ metallicity to find a residual offset from the polynomial fit. On a per-galaxy basis, we find the median value of this offset. If this median offset is 0, or close to it, then that galaxy's spaxels lie equally above and below our polynomial fit, with no systematic shift. 
 We iterate through all metallicity pairings, and assemble these median offsets for each galaxy, in each pairing. For each conversion, we then have a distribution of offsets for all of our galaxies.

 Given that the polynomial fit is itself closely tied to the median of all galaxies, we should not expect to see the peak of the distribution for all galaxies to be shifted away from zero. However, the breadth of the median offset distribution gives a sense of how much variability there is in the sample.  If the range in median offsets is near zero, but the overall $2\sigma$ scatter is large, then even for a single galaxy, the conversion between metallicity calibrations is subject to significant scatter, but between galaxies, there is not a lot of variability. By contrast, we may be summing over narrower distributions, but with a broad range in where these distributions lay in our parameter space; this scenario would be marked by a larger range in the median offsets from the polynomial fit. 

\begin{figure} 
\includegraphics[width=\columnwidth]{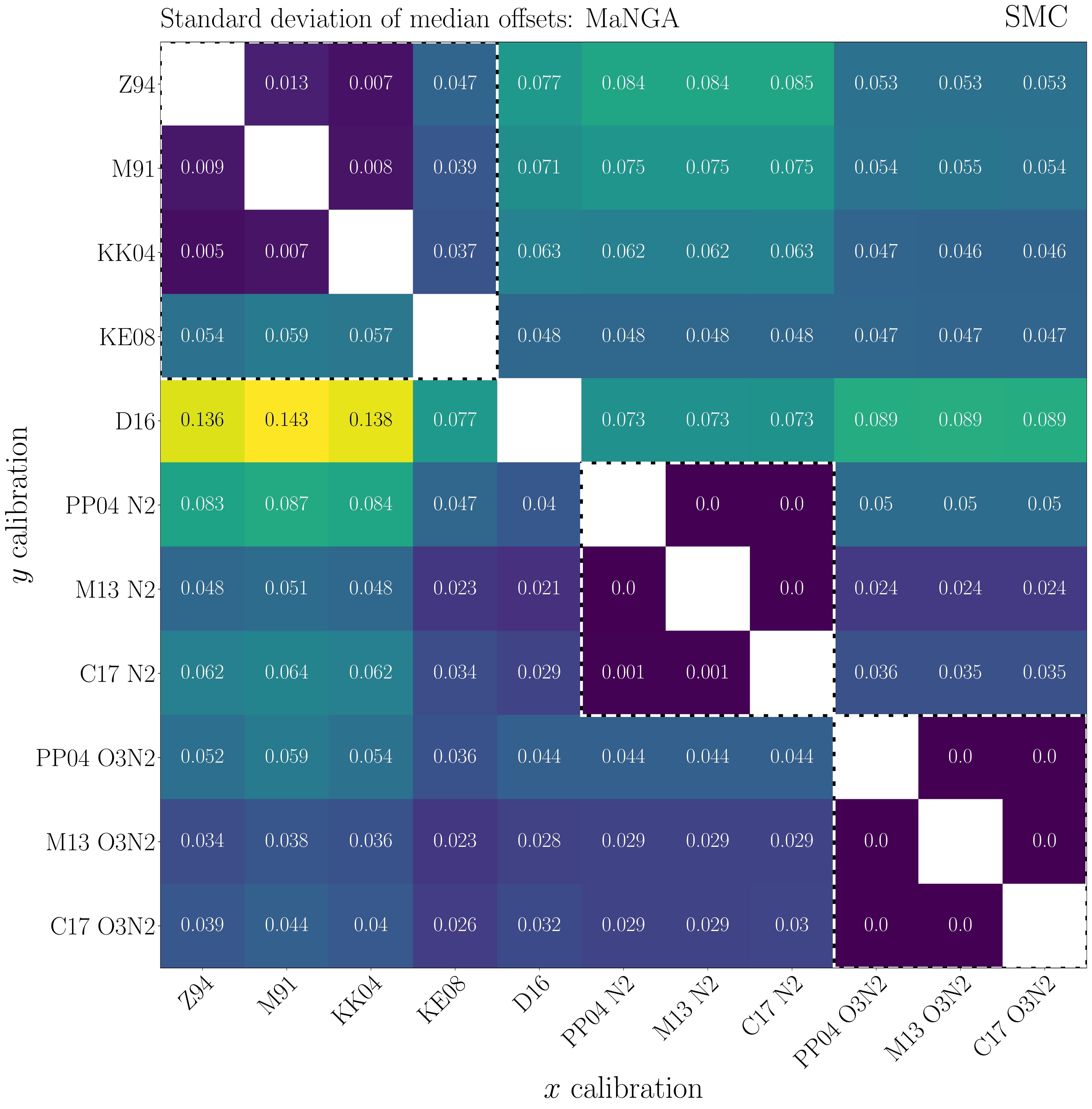}
  \caption{For the MaNGA sample, we show the standard deviation of the median offset for each galaxy's spaxels from the polynomial fit. This median represents the 4824 unique galaxies in MaNGA. For the set of calibrations with low overall scatter, we again find very small standard deviations in the median values, which indicates low galaxy-to-galaxy variability. On the other hand, larger standard deviations indicate that galaxies typically are much more variable in where they lie. R$_{23}$-based calibrations converting into D16 have particularly strong variability from galaxy to galaxy.}
  \label{fig:dr15_stdev_bygal}
\end{figure}

To quantify this, for each metallicity conversion, we take the set of median offsets for each galaxy, and find the standard deviation of the distribution. This should indicate the breadth of the distribution without being unduly sensitive to individual galaxy outliers. We summarize these standard deviations in Figure \ref{fig:dr15_stdev_bygal}, which is color coded by the standard deviation of the median offsets for our galaxy-by-galaxy assessment, and has the standard deviation, rounded to three decimal places, indicated in the center of each square. 

Typical standard deviation is 0.047 dex, with minima found for those calibrations with minimal $2\sigma$, and a maximum of 0.143 dex for M91 $\rightarrow$ D16. The overall $2\sigma$ scatter for that conversion is 0.264 dex, so this standard deviation could be responsible for up to $\sim$50\% of the total scatter. As with the R$_e$ test, we can test this directly. 

\begin{figure} 
\includegraphics[width=\columnwidth]{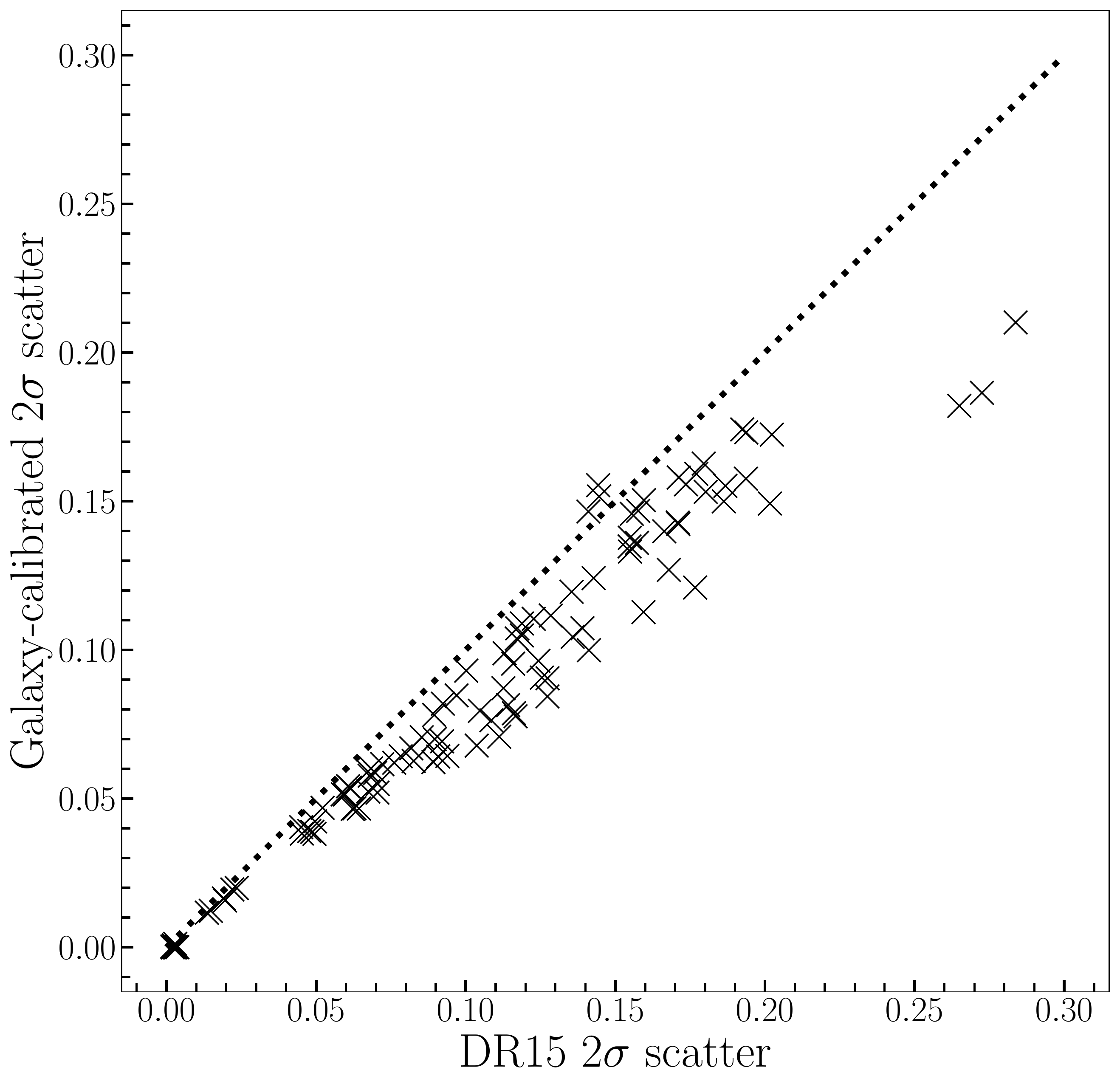}
  \caption{For the MaNGA sample, we show the reduction in scatter if the galaxy-to-galaxy variability is removed. On the vertical axis, we plot the $2\sigma$ scatter found after removing any median offset for each galaxy. On the horizontal axis we show the original, unmodified $2\sigma$ scatter for the MaNGA sample. In a diagonal black dotted line, we show the 1:1 relation, where the two match. The typical change to the scatter is of order 0.015 dex, though for the R$_{23}$ based calibrations into D16, we note that the reduction in scatter is much more substantial ($\sim$0.08 dex). Fractionally, the galaxy calibrated $2\sigma$ values are about 17 per cent lower than their corresponding uncalibrated points.}
  \label{fig:dr15_change_bygal}
\end{figure}

For each galaxy, we can subtract the median offset from the polynomial fit from all of its constituent spaxels. This should have the effect of manually shifting the galaxy's spaxels to be centered on the polynomial fit. As a result, if the galaxy-to-galaxy variability is a strong influence on the total scatter, we should see a general reduction in scatter. We show the results of this galaxy correction in Figure \ref{fig:dr15_change_bygal}. 
On the $y$ axis, we show the $2\sigma$ scatter after the galaxy-to-galaxy variability has been artificially removed, and on the $x$ axis, we show the $2\sigma$ values for the unmodified MaNGA sample. A 1-1 line is shown in a black dotted line. 
In general, the scatter around the polynomial fit is reduced by 0.015 dex as a result of this correction, with a maximum change of 0.09 dex (Z94 $\rightarrow$ D16). There are three cases where scatter increases (O3N2 $\rightarrow$ M91), but this increase is, fractionally, small. The impact (in dex) is strongest where the scatter was the highest; but fractionally, the reduction is relatively consistent across all calibrations. 
The typical reduction in the scatter as a result of removing individual galaxy variation is 17 per cent. However, we note that this is a median with a broad distribution; for individual conversions this can range from 5 per cent to 35 per cent of the total scatter. The median overall scatter in the MaNGA sample is $\sim0.1$ dex; after this correction, removing the impact of galaxy-to-galaxy variation, the $2\sigma$ scatter is reduced to $\sim0.08$ dex. 

We may speculate that galaxy-to-galaxy variability is present for all caibrations, but for those with the most emission lines in common, any bias in a change to the strength of the emission line will cancel out in the conversion, as any elevation or suppression in an emission line strength is present in both $x$ \& $y$ calibrations. However, for conversions with fewer emission lines in common (such as the R$_{23}$-based calibrations converting into D16), any emission lines which systematically change from galaxy to galaxy, or which are poorly predicted by the set of emission lines used by the $x$ calibration, may result in increased scatter. This may also be at least partially driving the strong polynomial offsets seen between D16 and the R$_{23}$-based calibrations in Figure \ref{fig:superfig_polyoffset}; as these calibration conversions do have strong galaxy to galaxy variability, and the TYPHOON data is only four galaxies, the small number statistics may be able to drive significant changes to the shape of a polynomial fit, where MaNGA and DR7 both have the benefit of increased galactic number statistics. 

\subsection{Emission line ratio correlations}

As the galaxy to galaxy variability does not fully account for the 2$\sigma$ scatter in the MaNGA sample, and as Figure \ref{fig:dr15_change_bygal} shows that even accounting for this variability leaves the high 2$\sigma$ scatter galaxies still at the highest values, there must be another strong factor present to drive scatter in the MaNGA relations. If, as proposed above, the remaining high scatter among certain calibrations is truly reflecting the relative weakness of an underlying correlation with emission line ratios, this weaker correlation should also appear statistically in this data. We thus perform a Spearman Rank correlation test.  We select all valid spaxels for each metallicity calibration, and all emission lines required for any of the other metallicity calibrations. We then, in all spaxels with high S/N emission line fluxes, calculate the emission line ratios N2, N2S2, O3N2, and R$_{23}$. We then perform a Spearman Rank correlation test between each metallicity calibration and all of the 4 possible line ratios, including the one upon which the metallicity is based. For the R$_{23}$ calibration, which is double-valued, we select only the upper branch (log(\NII/\OII) > --1.2) of the R$_{23}$ line ratio. The majority of the spaxels are found in the upper branch, and including the lower branch (where any correlation will invert) will weaken the intrinsic anti-correlation of the upper branch with metallicity. 

\begin{figure}
  \includegraphics[width=\columnwidth]{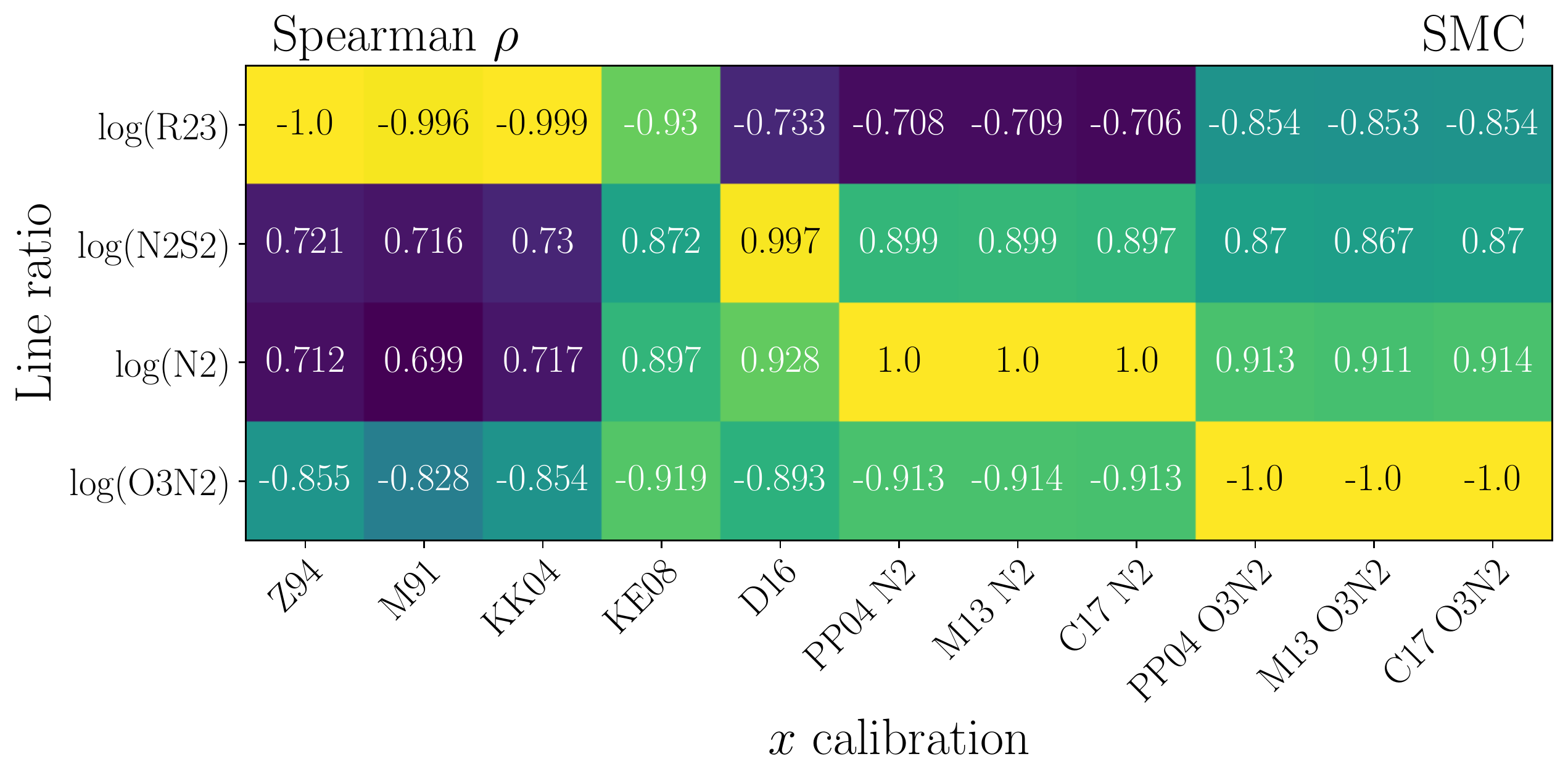}
  \caption{The results of the Spearman Rank correlation test between each metallicity calibration and the line ratios used among the full sample. The correlation coefficients are indicated in the center of each box, and the colorbar is mapped to the absolute value of the correlation coefficient. Metallicities are most strongly correlated with the emission line used for their own calibration, as expected. N2-based calibrations are most weakly correlated with the R$_{23}$ line ratio, as is D16. Conversely, the three oldest $R_{23}$ metallicity calibrations are most weakly correlated with the N2 and N2S2 line ratios.}
  \label{fig:spearman_matrix}
\end{figure}

The Spearman rank correlation tests for a nonparametric correlation between any two values, and produces a correlation coefficient ($\rho$) alongside a p-value which indicates the null hypothesis of uncorrelated data. All metallicities and all emission line ratios are able to exclude the null hypothesis of uncorrelated data (p-val = 0.0 in all cases). The correlation coefficient $\rho$ ranges from +1 (perfect positive correlation) to -1 (perfect negative correlation), with 0 indicating no correlation. The results of this test are shown in Figure \ref{fig:spearman_matrix}, where we show in colors the absolute value of the correlation coefficient $\rho$, and inside each square, we display $\rho$. All metallicity calibrations are extremely well correlated (by definition) to the emission line ratios which they make use of, color-coded in yellow.

The correlations are in general reasonable; no metallicity has a correlation with a line ratio used for another calibrator of less than ~0.7. However, it is the case that many of the calibrations correlate worse with certain line ratios than with others. The oldest R$_{23}$-based calibrations, M91, Z94, and KK04, for instance, have the weakest correlations with the N2 and N2S2 line ratios out of the entire sample; this matches their location at the highest $2\sigma$ scatter for those conversions in MaNGA. 

\begin{figure}
  \includegraphics[width='3in]{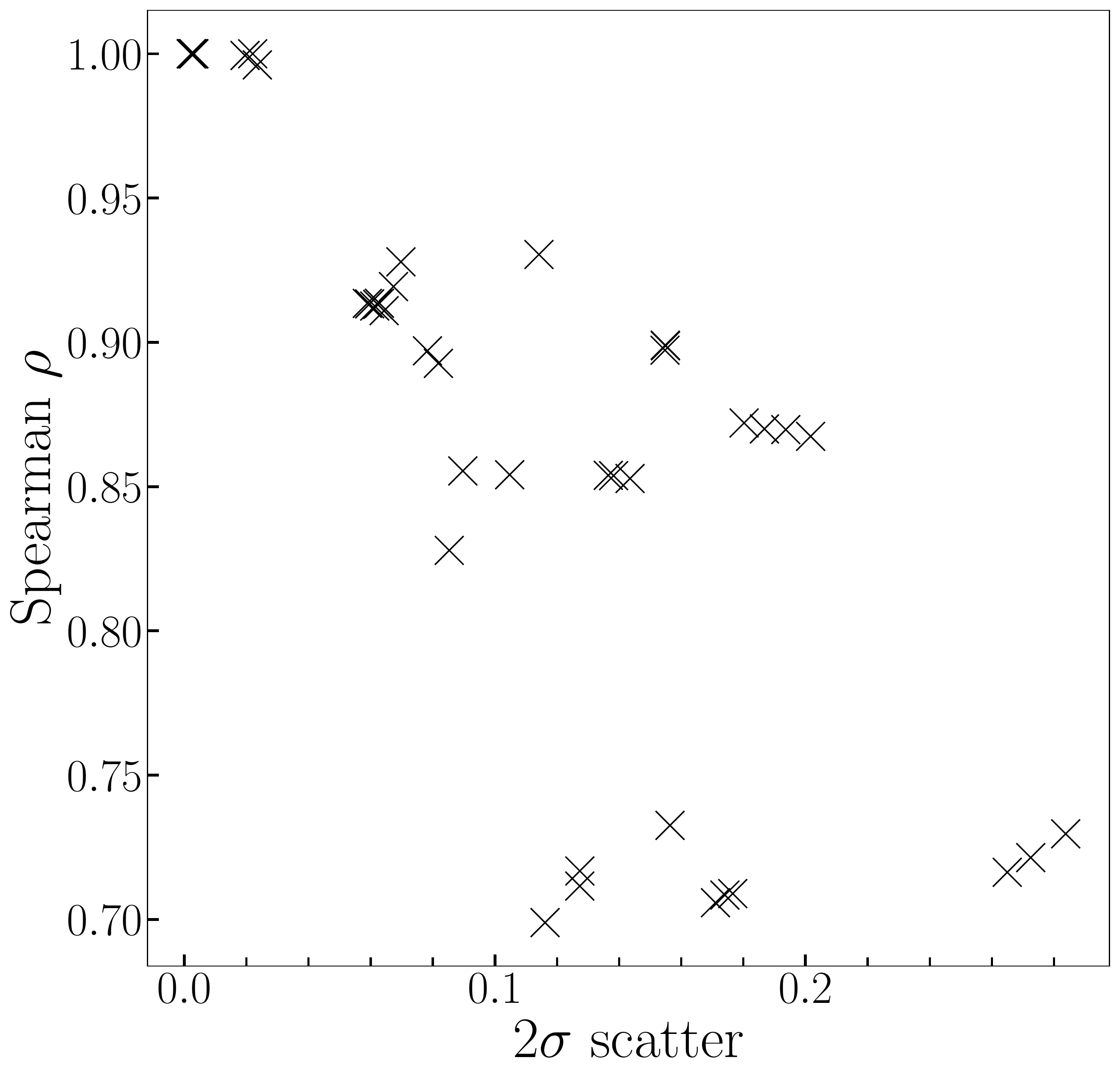}
  \caption{Comparison of the median 2$\sigma$ scatter for each metallicity converting into all calibrations which require a given emission line ratio, vs the Spearman Rank correlation test for the metallicity's correlation with that emission line ratio. Crosses at 2$\sigma$ scatter of 0.0 and correlations of 1.0 indicate the metallicity's correlation with its own line ratio.  }
  \label{fig:spearman_scatter}
\end{figure}

In fact, if we compare ascending 2$\sigma$ scatter in the MaNGA sample with the increasing strength of the correlation between calibration and emission line ratios, we find a good match, illustrated in Figure \ref{fig:spearman_scatter}. The highest scatters are exclusively found with $x$ metallicity calibrations less well correlated with the emission line ratios required of the $y$ metallicity. Lowest scatter conversions are found among those with stronger correlations between $x$ calibration and $y$ line ratio.

We conclude, as a result of these final tests, that while calibrations which use consistent emission lines have minimal scatter in the first place, and are not substantially impacted by any further biases, calibrations which use different emission lines are more sensitive to variations on a galaxy-to-galaxy basis, and are subject to increased scatter even after this is accounted for. However, the radial location of a given spaxel does not appear to cause a significant shift; correcting for effective radius typically only decreases the scatter by 0.0006 dex. By contrast, corrections for galaxy-to-galaxy variations reduce the scatter by 0.015 dex, accounting for (typically) 17 per cent of the total $2\sigma$ scatter, though for extreme cases, the variation is 0.08 dex and 35 per cent of the total scatter. Even after this correction is undertaken, those conversions which rely on different emission lines remain with higher scatter, as the galaxy-to-galaxy variability broadly represents a consistent fraction of the overall scatter. This remaining higher scatter is likely due to the relative weaker correlation between certain metallicity calibrations and the emission line ratios used by other metrics. 

These two issues, increased galaxy-to-galaxy variability, and poor underlying correlations with emission line ratios used for other calibrations, can naturally coexist. Each galaxy may be subject to different physical conditions, and the metallicity calibrations used here are sensitive to such differences in physical characteristics, depending on their theoretical underpinnings and/or emission line ratios. Since emission line strengths are not perfectly predictive of each other, we can imagine that converting between metallicity calibrations relying on different emission line ratios will indeed have broader or narrower distributions. Broadened scatter can thus be a reflection of both increased variability from galaxy to galaxy, and that specific emission line ratios are not perfectly predictive of other line ratios. 

The emission lines of different galaxies may replicate the effect seen by T21 when transitioning from the DR4 to the DR7. As emission line strengths changed between the two SDSS data releases, this change introduced a systematic bias for galaxies whose metallicities relied on different emission lines. For galaxies where metallicity metrics were based upon the same line ratios, the two data releases were very consistent, as this enhancement in the emission line strength was present in both $x$ \& $y$ calibrations, and so in the conversion, there was no systematic effect. In the current work, while we are instead looking at differences between individual galaxies, rather than a reprocessing of spectral data, this will reflect variability in the star forming regions in different galaxies generating slightly different emission line ratios.

While it is beyond the scope of the current work to provide a comprehensive study of why certain emission line ratios would correlate less well with each other, or have increased variability from galaxy to galaxy, we do note that these metal lines are produced in different physical environments. In principle, we should expect that the line ratios using metal lines produced in similar physical environments should correlate better between them. \citet{Kewley2019} presented predictions from {\sc{mappings}} 5.1 to indicate which regions of a nebula surrounding a photoionizing source are responsible for different emission line fluxes (their Figs. 2 \& 3). \OIII~is produced throughout, though stronger closer to the ionizing source. \NII, along with \OII, is also found throughout the nebula, but with a stronger preference for emission further from the ionizing source. \SII, by contrast, is strongly concentrated far from the ionizing source, in partially ionized gas. Compounding these issues are differences in theoretical models underpinning the conversions from line ratio to metallicity, or differences in data sets for which ``direct'' measurements are possible. We therefore present these correlations by way of informing the reader as to which conversions are more or less likely to encounter significant scatter, and allow the user to determine whether this scatter and galaxy variability will impede on their own science questions. 

\section{Conclusions}
\label{sec:conclusions}

We have computed and make available a set of 5th order polynomial fits, allowing for the conversion between 11 different metallicity calibrations, based on the public MaNGA DR15 data release, for both SMC and MW type dust corrections. Each fit is based on $\sim$ 1.1 million spaxels, and represents the largest sample ever considered for conversions between metallicity calibrations. We analyse the resultant conversions between calibrations for trends and assess our sample for biases in covering fraction, DIG contamination, and conclude that the largest influence over which conversions in MaNGA are subject to increased scatter is simply that of relying on different sets of emission lines. 

The results of this work can be summarized as:

\begin{itemize}
  \item Polynomial fits are repeated for both SMC and MW dust correction curves. In general we do not find a large discrepancy between the two dust correction models; however, the SMC curve typically has a slightly larger number of valid spaxels due to a change in the distribution of spaxels in the BPT diagram, and the scatter around the polynomial is typically slightly larger for the MW curve. The figures in this work show the SMC dust correction curve data; the MW dust corrected metallicity figures are available online. 
 \item Conversions between calibrations are insensitive to redshift binning over the range of redshifts present in the MaNGA data set ($ 0 < z < 0.15$). This is consistent with no substantive dependence on covering fraction.
 \item 2$\sigma$ scatter in MaNGA around the polynomial fits ranges from 0.002 dex to 0.287 dex, with a typical value of 0.1 dex. Calibrations with fewer emission lines in common are broadly subject to increased scatter in MaNGA, and especially so for calibrations which only share one emission line. Metallicity calibration pairs with a larger fraction of lines in overlap typically display smaller scatter.  
   \item The effective radius of a given spaxel does not show a strong trend with the offset of a spaxel from the polynomial fit, and is not a driving factor of the scatter around a given polynomial fit.
\item We test for the influence of DIG light, and determine that its influence in the current work should be minimal, as our combination of S/N, EW cuts, and BPT classifications have excluded the majority of DIG-dominated spaxels from our sample. The spaxels in this samples used in this work are $> 99$ per cent HII region-dominated.
  \item There are significant galaxy-to-galaxy systematic offsets from the polynomial fit; these offsets are strongest in situations where emission lines are less in overlap, and account for 5 - 35 per cent (typically 17 per cent) of the $2\sigma$ scatter present in the MaNGA sample. These galaxy-to-galaxy variations partially account for the very different levels of scatter in the MaNGA sample from calibration to calibration. 
  \item The polynomial fits for the MaNGA data match comparison data sets at different resolutions (the SDSS DR7 \& the IFS survey TYPHOON), and do so particularly well in cases where the scatter in the conversion between calibrations in MaNGA is very small, which is also where galaxy-to-galaxy variability is smallest. This occurs preferentially when the emission lines required for both calibrations are consistent. The median offset between polynomials fit to different data sets is 0.018 dex. We conclude that our conversions are viable over a broad range of local redshifts (0.001 $< z < $ 0.15). 
   \end{itemize}

We do not intend this work to indicate a ``preferred'' metallicity calibration, but to allow conversions between calibrations for more direct comparisons between works using different metallicity calibrations in the MaNGA data set and beyond, and to make the user aware of potential issues when converting between different metallicity calibrations.

Due to the large data volume presented here, the figures in this manuscript, additional figures, and full, machine-readable, versions of the table in the Appendix, along with a parallel set of the same for the MW dust correction curve, are available on line as supplementary material. 
\section*{Acknowledgements}

We thank the referee for a constructive review, which has improved the clarity of this manuscript.

JMS thanks medical professionals for their work during this COVID-19 pandemic, without whom this work would not have been feasible to complete over the last year.  SLE gratefully acknowledges the receipt of an NSERC Discovery grant.

 Figures in this manuscript were generated using the {\sc{matplotlib}} python plotting package \citep{matplotlib}. Analysis made use of the {\sc{scipy}} package of statistical methods \citep{scipy}, and the {\sc{numpy}} package for array-type data handling \citep{numpy, numpy2020}.  
We also made use of the cosmology calculator of \citet{Wright2006}.

Funding for the Sloan Digital Sky Survey IV has been provided by the Alfred P. Sloan Foundation, the U.S. Department of Energy Office of Science, and the Participating Institutions. SDSS-IV acknowledges
support and resources from the Center for High-Performance Computing at
the University of Utah. The SDSS web site is www.sdss.org.

SDSS-IV is managed by the Astrophysical Research Consortium for the 
Participating Institutions of the SDSS Collaboration including the 
Brazilian Participation Group, the Carnegie Institution for Science, 
Carnegie Mellon University, the Chilean Participation Group, the French Participation Group, Harvard-Smithsonian Center for Astrophysics, 
Instituto de Astrof\'isica de Canarias, The Johns Hopkins University, Kavli Institute for the Physics and Mathematics of the Universe (IPMU) / 
University of Tokyo, the Korean Participation Group, Lawrence Berkeley National Laboratory, 
Leibniz Institut f\"ur Astrophysik Potsdam (AIP),  
Max-Planck-Institut f\"ur Astronomie (MPIA Heidelberg), 
Max-Planck-Institut f\"ur Astrophysik (MPA Garching), 
Max-Planck-Institut f\"ur Extraterrestrische Physik (MPE), 
National Astronomical Observatories of China, New Mexico State University, 
New York University, University of Notre Dame, 
Observat\'ario Nacional / MCTI, The Ohio State University, 
Pennsylvania State University, Shanghai Astronomical Observatory, 
United Kingdom Participation Group,
Universidad Nacional Aut\'onoma de M\'exico, University of Arizona, 
University of Colorado Boulder, University of Oxford, University of Portsmouth, 
University of Utah, University of Virginia, University of Washington, University of Wisconsin, 
Vanderbilt University, and Yale University.

\section*{Data availability statement}
The emission line data underlying this manuscript are publicly available as part of the MaNGA DR15 data release, available at \href{https://www.sdss.org/dr15/}{https://www.sdss.org/dr15/}. Polynomial fits are available in the online supplementary material. Metallicity values themselves are available upon reasonable request to the corresponding author. 




\bibliographystyle{mnras}
\bibliography{library} 

\begin{thebibliography}{}
\makeatletter
\relax
\def\mn@urlcharsother{\let\do\@makeother \do\$\do\&\do\#\do\^\do\_\do\%\do\~}
\def\mn@doi{\begingroup\mn@urlcharsother \@ifnextchar [ {\mn@doi@}
  {\mn@doi@[]}}
\def\mn@doi@[#1]#2{\def\@tempa{#1}\ifx\@tempa\@empty \href
  {http://dx.doi.org/#2} {doi:#2}\else \href {http://dx.doi.org/#2} {#1}\fi
  \endgroup}
\def\mn@eprint#1#2{\mn@eprint@#1:#2::\@nil}
\def\mn@eprint@arXiv#1{\href {http://arxiv.org/abs/#1} {{\tt arXiv:#1}}}
\def\mn@eprint@dblp#1{\href {http://dblp.uni-trier.de/rec/bibtex/#1.xml}
  {dblp:#1}}
\def\mn@eprint@#1:#2:#3:#4\@nil{\def\@tempa {#1}\def\@tempb {#2}\def\@tempc
  {#3}\ifx \@tempc \@empty \let \@tempc \@tempb \let \@tempb \@tempa \fi \ifx
  \@tempb \@empty \def\@tempb {arXiv}\fi \@ifundefined
  {mn@eprint@\@tempb}{\@tempb:\@tempc}{\expandafter \expandafter \csname
  mn@eprint@\@tempb\endcsname \expandafter{\@tempc}}}

\bibitem[\protect\citeauthoryear{Abazajian et~al.,}{Abazajian
  et~al.}{2009}]{DR7}
Abazajian K.~N.,  et~al., 2009, \mn@doi [\apjs] {10.1088/0067-0049/182/2/543},
  182, 543

\bibitem[\protect\citeauthoryear{Adelman-McCarthy et~al.,}{Adelman-McCarthy
  et~al.}{2006}]{DR4}
Adelman-McCarthy J.~K.,  et~al., 2006, \mn@doi [\apjs] {10.1086/497917}, 162,
  38

\bibitem[\protect\citeauthoryear{Aguado et~al.,}{Aguado
  et~al.}{2019}]{Aguado2019}
Aguado D.~S.,  et~al., 2019, \mn@doi [\apjs] {10.3847/1538-4365/aaf651}, 240,
  23

\bibitem[\protect\citeauthoryear{Alloin, Collin-Souffrin, Joly  \&
  Vigroux}{Alloin et~al.}{1979}]{Alloin1979}
Alloin D.,  Collin-Souffrin S.,  Joly M.,   Vigroux L.,  1979, \aap, 78, 200

\bibitem[\protect\citeauthoryear{Andrews \& Martini}{Andrews \&
  Martini}{2013}]{Andrews2013}
Andrews B.~H.,  Martini P.,  2013, \mn@doi [\apj]
  {10.1088/0004-637X/765/2/140}, 765, 140

\bibitem[\protect\citeauthoryear{Baldwin, Phillips  \& Terlevich}{Baldwin
  et~al.}{1981}]{Baldwin1981}
Baldwin J.~A.,  Phillips M.~M.,   Terlevich R.,  1981, \mn@doi [\pasp]
  {10.1086/130766}, 93, 5

\bibitem[\protect\citeauthoryear{Barrera-Ballesteros
  et~al.,}{Barrera-Ballesteros et~al.}{2015}]{Barrera-Ballesteros2015}
Barrera-Ballesteros J.~K.,  et~al., 2015, \mn@doi [\aap]
  {10.1051/0004-6361/201425397}, 579, A45

\bibitem[\protect\citeauthoryear{Belfiore et~al.,}{Belfiore
  et~al.}{2015}]{Belfiore2014}
Belfiore F.,  et~al., 2015, \mn@doi [\mnras] {10.1093/mnras/stv296}, 449, 867

\bibitem[\protect\citeauthoryear{Belfiore et~al.,}{Belfiore
  et~al.}{2017}]{Belfiore2017}
Belfiore F.,  et~al., 2017, \mn@doi [\mnras] {10.1093/mnras/stx789}, 469, 151

\bibitem[\protect\citeauthoryear{Bian, Kewley  \& Dopita}{Bian
  et~al.}{2018}]{Bian2018}
Bian F.,  Kewley L.~J.,   Dopita M.~A.,  2018, \mn@doi [\apj]
  {10.3847/1538-4357/aabd74}, 859, 175

\bibitem[\protect\citeauthoryear{Blanton et~al.,}{Blanton
  et~al.}{2017}]{Blanton2017}
Blanton M.~R.,  et~al., 2017, \mn@doi [\aj] {10.3847/1538-3881/aa7567}, 154, 28

\bibitem[\protect\citeauthoryear{Brown, Martini  \& Andrews}{Brown
  et~al.}{2016}]{Brown2016}
Brown J.~S.,  Martini P.,   Andrews B.~H.,  2016, \mn@doi [\mnras]
  {10.1093/mnras/stw392}, 458, 1529

\bibitem[\protect\citeauthoryear{Bundy et~al.,}{Bundy et~al.}{2015}]{Bundy2015}
Bundy K.,  et~al., 2015, \mn@doi [\apj] {10.1088/0004-637X/798/1/7}, 798, 7

\bibitem[\protect\citeauthoryear{Bustamante, Sparre, Springel  \&
  Grand}{Bustamante et~al.}{2018}]{Bustamente2018}
Bustamante S.,  Sparre M.,  Springel V.,   Grand R. J.~J.,  2018, \mn@doi
  [\mnras] {10.1093/mnras/sty1692}, 479, 3381

\bibitem[\protect\citeauthoryear{Cardelli, Clayton  \& Mathis}{Cardelli
  et~al.}{1989}]{ccm}
Cardelli J.~A.,  Clayton G.~C.,   Mathis J.~S.,  1989, \mn@doi [\apj]
  {10.1086/167900}, 345, 245

\bibitem[\protect\citeauthoryear{Cid~Fernandes, Stasi{\'{n}}ska, Mateus  \&
  Vale-Asari}{Cid~Fernandes et~al.}{2011}]{Fernandes2011}
Cid~Fernandes R.,  Stasi{\'{n}}ska G.,  Mateus A.,   Vale-Asari N.,  2011,
  \mn@doi [\mnras] {10.1111/j.1365-2966.2011.18244.x}, 413, 1687

\bibitem[\protect\citeauthoryear{Curti, Cresci, Mannucci, Marconi, Maiolino  \&
  Esposito}{Curti et~al.}{2017}]{Curti2017}
Curti M.,  Cresci G.,  Mannucci F.,  Marconi A.,  Maiolino R.,   Esposito S.,
  2017, \mn@doi [\mnras] {10.1093/mnras/stw2766}, 465, 1384

\bibitem[\protect\citeauthoryear{Curti, Mannucci, Cresci  \& Maiolino}{Curti
  et~al.}{2020}]{Curti2020}
Curti M.,  Mannucci F.,  Cresci G.,   Maiolino R.,  2020, \mn@doi [\mnras]
  {10.1093/mnras/stz2910}, 491, 944

\bibitem[\protect\citeauthoryear{Denicol{\'o}, Terlevich  \&
  Terlevich}{Denicol{\'o} et~al.}{2002}]{Denicolo2002}
Denicol{\'o} G.,  Terlevich R.,   Terlevich E.,  2002, \mn@doi [\mnras]
  {10.1046/j.1365-8711.2002.05041.x}, 330, 69

\bibitem[\protect\citeauthoryear{Dopita, Kewley, Sutherland  \&
  Nicholls}{Dopita et~al.}{2016}]{Dopita2016}
Dopita M.~A.,  Kewley L.~J.,  Sutherland R.~S.,   Nicholls D.~C.,  2016,
  \mn@doi [\apss] {10.1007/s10509-016-2657-8}, 361, 61

\bibitem[\protect\citeauthoryear{Drory et~al.,}{Drory et~al.}{2015}]{Drory2015}
Drory N.,  et~al., 2015, \mn@doi [\aj] {10.1088/0004-6256/149/2/77}, 149, 77

\bibitem[\protect\citeauthoryear{Ellison, Patton, Simard  \&
  McConnachie}{Ellison et~al.}{2008}]{Ellison2008}
Ellison S.~L.,  Patton D.~R.,  Simard L.,   McConnachie A.~W.,  2008, \mn@doi
  [arXiv] {10.1088/0004-6256/135/5/1877}, pp 1877--1899

\bibitem[\protect\citeauthoryear{Ferland, Korista, Verner, Ferguson, Kingdon
  \& Verner}{Ferland et~al.}{1998}]{Ferland1998}
Ferland G.~J.,  Korista K.~T.,  Verner D.~A.,  Ferguson J.~W.,  Kingdon J.~B.,
   Verner E.~M.,  1998, \mn@doi [\pasp] {10.1086/316190}, 110, 761

\bibitem[\protect\citeauthoryear{Fioc \& Rocca-Volmerange}{Fioc \&
  Rocca-Volmerange}{1997}]{Fioc1997}
Fioc M.,  Rocca-Volmerange B.,  1997, \aap, 326, 950

\bibitem[\protect\citeauthoryear{Groves, Dopita  \& Sutherland}{Groves
  et~al.}{2004a}]{Groves2004a}
Groves B.~A.,  Dopita M.~A.,   Sutherland R.~S.,  2004a, \mn@doi [\apjs]
  {10.1086/421113}, 153, 9

\bibitem[\protect\citeauthoryear{Groves, Dopita  \& Sutherland}{Groves
  et~al.}{2004b}]{Groves2004b}
Groves B.~A.,  Dopita M.~A.,   Sutherland R.~S.,  2004b, \mn@doi [\apjs]
  {10.1086/421114}, 153, 75

\bibitem[\protect\citeauthoryear{Gunn, Siegmund  \& al}{Gunn
  et~al.}{2006}]{Gunn2006}
Gunn J.~E.,  Siegmund W.~A.,   al E. J. M.~e.,  2006, \mn@doi [arXiv]
  {10.1086/500975}, p.~2332

\bibitem[\protect\citeauthoryear{Haffner et~al.,}{Haffner
  et~al.}{2009}]{Haffner2009}
Haffner L.~M.,  et~al., 2009, \mn@doi [Reviews of Modern Physics]
  {10.1103/RevModPhys.81.969}, 81, 969

\bibitem[\protect\citeauthoryear{Harris et~al.,}{Harris
  et~al.}{2020}]{numpy2020}
Harris C.~R.,  et~al., 2020, \mn@doi [\nat] {10.1038/s41586-020-2649-2}, 585,
  357

\bibitem[\protect\citeauthoryear{Hunter}{Hunter}{2007}]{matplotlib}
Hunter J.~D.,  2007, \mn@doi [Computing in Science and Engineering]
  {10.1109/MCSE.2007.55}, 9, 90

\bibitem[\protect\citeauthoryear{Kauffmann et~al.,}{Kauffmann
  et~al.}{2003a}]{Kauffmann2003a}
Kauffmann G.,  et~al., 2003a, \mn@doi [\mnras]
  {10.1046/j.1365-8711.2003.06291.x}, 341, 33

\bibitem[\protect\citeauthoryear{Kauffmann et~al.,}{Kauffmann
  et~al.}{2003b}]{Kauffmann2003}
Kauffmann G.,  et~al., 2003b, \mn@doi [\mnras]
  {10.1111/j.1365-2966.2003.07154.x}, 346, 1055

\bibitem[\protect\citeauthoryear{Kewley \& Dopita}{Kewley \&
  Dopita}{2002}]{Kewley2002}
Kewley L.~J.,  Dopita M.~A.,  2002, \mn@doi [\apjs] {10.1086/341326}, 142, 35

\bibitem[\protect\citeauthoryear{Kewley \& Ellison}{Kewley \&
  Ellison}{2008}]{Kewley2008}
Kewley L.~J.,  Ellison S.~L.,  2008, \mn@doi [\apj] {10.1086/587500}, 681, 1183

\bibitem[\protect\citeauthoryear{Kewley, Rupke, Jabran~Zahid, Geller  \&
  Barton}{Kewley et~al.}{2010}]{Kewley2010a}
Kewley L.~J.,  Rupke D.,  Jabran~Zahid H.,  Geller M.~J.,   Barton E.~J.,
  2010, \mn@doi [\apj] {10.1088/2041-8205/721/1/L48}, 721, L48

\bibitem[\protect\citeauthoryear{Kewley, Nicholls  \& Sutherland}{Kewley
  et~al.}{2019}]{Kewley2019}
Kewley L.~J.,  Nicholls D.~C.,   Sutherland R.~S.,  2019, \mn@doi [\araa]
  {10.1146/annurev-astro-081817-051832}, 57, 511

\bibitem[\protect\citeauthoryear{Kobulnicky \& Kewley}{Kobulnicky \&
  Kewley}{2004}]{KK04}
Kobulnicky H.~A.,  Kewley L.~J.,  2004, \mn@doi [\apj] {10.1086/425299}, 617,
  240

\bibitem[\protect\citeauthoryear{Kobulnicky et~al.,}{Kobulnicky
  et~al.}{2003}]{Kobulnicky2003}
Kobulnicky H.~A.,  et~al., 2003, \mn@doi [\apj] {10.1086/379360}, 599, 1006

\bibitem[\protect\citeauthoryear{Kreckel et~al.,}{Kreckel
  et~al.}{2020}]{Kreckel2020}
Kreckel K.,  et~al., 2020, \mn@doi [arXiv] {10.1093/mnras/staa2743}, p.
  arXiv:2009.02342

\bibitem[\protect\citeauthoryear{Lara-Lopez et~al.,}{Lara-Lopez
  et~al.}{2010}]{Lara-Lopez2010}
Lara-Lopez M.~A.,  et~al., 2010, \mn@doi [\aap] {10.1051/0004-6361/201014803},
  521, L53

\bibitem[\protect\citeauthoryear{Law et~al.,}{Law et~al.}{2015}]{Law2015}
Law D.~R.,  et~al., 2015, ] {10.1088/0004-6256/150/1/19}, 150, 19

\bibitem[\protect\citeauthoryear{Law et~al.,}{Law et~al.}{2016}]{Law2016}
Law D.~R.,  et~al., 2016, \mn@doi [\aj] {10.3847/0004-6256/152/4/83}, 152, 83

\bibitem[\protect\citeauthoryear{Law et~al.,}{Law et~al.}{2020}]{Law2020}
Law D.~R.,  et~al., 2020, arXiv, p. arXiv:2011.06012

\bibitem[\protect\citeauthoryear{Lee, Skillman, Cannon, Jackson, Gehrz,
  Polomski  \& Woodward}{Lee et~al.}{2006}]{Lee2006}
Lee H.,  Skillman E.~D.,  Cannon J.~M.,  Jackson D.~C.,  Gehrz R.~D.,  Polomski
  E.~F.,   Woodward C.~E.,  2006, \mn@doi [\apj] {10.1086/505573}, 647, 970

\bibitem[\protect\citeauthoryear{Leitherer et~al.,}{Leitherer
  et~al.}{1999}]{Leitherer1999}
Leitherer C.,  et~al., 1999, \mn@doi [\apjs] {10.1086/313233}, 123, 3

\bibitem[\protect\citeauthoryear{Lequeux, Peimbert, Rayo, Serrano  \&
  Torres-Peimbert}{Lequeux et~al.}{1979}]{Lequeux1979}
Lequeux J.,  Peimbert M.,  Rayo J.~F.,  Serrano A.,   Torres-Peimbert S.,
  1979, \aap, 80, 155

\bibitem[\protect\citeauthoryear{Li, Krumholz, Wisnioski, Mendel, Kewley,
  S{\'a}nchez  \& Galbany}{Li et~al.}{2021}]{Li2021}
Li Z.,  Krumholz M.~R.,  Wisnioski E.,  Mendel J.~T.,  Kewley L.~J.,
  S{\'a}nchez S.~F.,   Galbany L.,  2021, arXiv, p. arXiv:2104.14807

\bibitem[\protect\citeauthoryear{Liang, Hammer  \& Flores}{Liang
  et~al.}{2006}]{Liang2006}
Liang Y.~C.,  Hammer F.,   Flores H.,  2006, \mn@doi [\aap]
  {10.1051/0004-6361:20053821}, 447, 113

\bibitem[\protect\citeauthoryear{Liang, Hammer, Yin, Flores, Rodrigues  \&
  Yang}{Liang et~al.}{2007}]{Liang2007}
Liang Y.~C.,  Hammer F.,  Yin S.~Y.,  Flores H.,  Rodrigues M.,   Yang Y.~B.,
  2007, \mn@doi [\aap] {10.1051/0004-6361:20077436}, 473, 411

\bibitem[\protect\citeauthoryear{Maier, Ziegler, Lilly, Contini, Perez~Montero,
  Lamareille, Bolzonella  \& Le~Floc'h}{Maier et~al.}{2015}]{Maier2014a}
Maier C.,  Ziegler B.~L.,  Lilly S.~J.,  Contini T.,  Perez~Montero E.,
  Lamareille F.,  Bolzonella M.,   Le~Floc'h E.,  2015, \mn@doi [\aap]
  {10.1051/0004-6361/201425224}, 577, A14

\bibitem[\protect\citeauthoryear{Maiolino et~al.,}{Maiolino
  et~al.}{2008}]{Maiolino2008}
Maiolino R.,  et~al., 2008, \mn@doi [\aap] {10.1051/0004-6361:200809678}, 488,
  463

\bibitem[\protect\citeauthoryear{Mannucci, Cresci, Maiolino, Marconi  \&
  Gnerucci}{Mannucci et~al.}{2010}]{Mannucci2010}
Mannucci F.,  Cresci G.,  Maiolino R.,  Marconi A.,   Gnerucci A.,  2010,
  \mn@doi [\mnras] {10.1111/j.1365-2966.2010.17291.x}, 408, 2115

\bibitem[\protect\citeauthoryear{Marino et~al.,}{Marino
  et~al.}{2013}]{Marino2013}
Marino R.~A.,  et~al., 2013, \mn@doi [\aap] {10.1051/0004-6361/201321956}, 559,
  A114

\bibitem[\protect\citeauthoryear{McGaugh}{McGaugh}{1991}]{m91}
McGaugh S.~S.,  1991, \mn@doi [\apj] {10.1086/170569}, 380, 140

\bibitem[\protect\citeauthoryear{Mihos \& Hernquist}{Mihos \&
  Hernquist}{1996}]{Mihos1996}
Mihos J.~C.,  Hernquist L.,  1996, \mn@doi [\apj] {10.1086/177353}, 464, 641

\bibitem[\protect\citeauthoryear{Pagel, Edmunds, Blackwell, Chun  \&
  Smith}{Pagel et~al.}{1979}]{Pagel1979}
Pagel B. E.~J.,  Edmunds M.~G.,  Blackwell D.~E.,  Chun M.~S.,   Smith G.,
  1979, \mn@doi [\mnras] {10.1093/mnras/189.1.95}, 189, 95

\bibitem[\protect\citeauthoryear{Pei}{Pei}{1992}]{Pei1992}
Pei Y.~C.,  1992, \mn@doi [\apj] {10.1086/171637}, 395, 130

\bibitem[\protect\citeauthoryear{Pettini \& Pagel}{Pettini \&
  Pagel}{2004}]{PP04}
Pettini M.,  Pagel B. E.~J.,  2004, \mn@doi [\mnras]
  {10.1111/j.1365-2966.2004.07591.x}, 348, L59

\bibitem[\protect\citeauthoryear{Pilyugin, Grebel, Zinchenko, Nefedyev, Shulga,
  Wei  \& Berczik}{Pilyugin et~al.}{2018}]{Pilyugin2018}
Pilyugin L.~S.,  Grebel E.~K.,  Zinchenko I.~A.,  Nefedyev Y.~A.,  Shulga
  V.~M.,  Wei H.,   Berczik P.~P.,  2018, \mn@doi [\aap]
  {10.1051/0004-6361/201732185}, 613, A1

\bibitem[\protect\citeauthoryear{Poetrodjojo, D'Agostino, Groves, Kewley, Ho,
  Rich, Madore  \& Seibert}{Poetrodjojo et~al.}{2019}]{Poetrodjojo2019}
Poetrodjojo H.,  D'Agostino J.~J.,  Groves B.,  Kewley L.,  Ho I.-T.,  Rich J.,
   Madore B.~F.,   Seibert M.,  2019, \mn@doi [\mnras] {10.1093/mnras/stz1241},
  487, 79

\bibitem[\protect\citeauthoryear{Reynolds}{Reynolds}{1984}]{Reynolds1984}
Reynolds R.~J.,  1984, \mn@doi [\apj] {10.1086/162190}, 282, 191

\bibitem[\protect\citeauthoryear{Reynolds}{Reynolds}{1990}]{Reynolds1990}
Reynolds R.~J.,  1990, \mn@doi [\apj] {10.1086/185640}, 349, L17

\bibitem[\protect\citeauthoryear{Rich, Torrey, Kewley, Dopita  \& Rupke}{Rich
  et~al.}{2012}]{Rich2012}
Rich J.~A.,  Torrey P.,  Kewley L.~J.,  Dopita M.~A.,   Rupke D. S.~N.,  2012,
  \mn@doi [\apj] {10.1088/0004-637X/753/1/5}, 753, 5

\bibitem[\protect\citeauthoryear{Rupke, Kewley  \& Barnes}{Rupke
  et~al.}{2010a}]{Rupke2010a}
Rupke D. S.~N.,  Kewley L.~J.,   Barnes J.~E.,  2010a, \mn@doi [\apj]
  {10.1088/2041-8205/710/2/L156}, 710, L156

\bibitem[\protect\citeauthoryear{Rupke, Kewley  \& Chien}{Rupke
  et~al.}{2010b}]{Rupke2010b}
Rupke D. S.~N.,  Kewley L.~J.,   Chien L.~H.,  2010b, \mn@doi [\apj]
  {10.1088/0004-637X/723/2/1255}, 723, 1255

\bibitem[\protect\citeauthoryear{S{\'a}nchez-Menguiano
  et~al.,}{S{\'a}nchez-Menguiano et~al.}{2016}]{Sanchez-Menguiano2016}
S{\'a}nchez-Menguiano L.,  et~al., 2016, \mn@doi [\aap]
  {10.1051/0004-6361/201527450}, 587, A70

\bibitem[\protect\citeauthoryear{Sanchez et~al.,}{Sanchez
  et~al.}{2014}]{Sanchez2014}
Sanchez S.~F.,  et~al., 2014, \mn@doi [\aap] {10.1051/0004-6361/201322343},
  563, A49

\bibitem[\protect\citeauthoryear{Sanchez et~al.,}{Sanchez
  et~al.}{2015}]{Sanchez2014a}
Sanchez S.~F.,  et~al., 2015, \mn@doi [\aap] {10.1051/0004-6361/201424873},
  574, A47

\bibitem[\protect\citeauthoryear{Sanchez et~al.,}{Sanchez
  et~al.}{2016a}]{Sanchez2016}
Sanchez S.~F.,  et~al., 2016a, \rmxaa, 52, 21

\bibitem[\protect\citeauthoryear{Sanchez et~al.,}{Sanchez
  et~al.}{2016b}]{Sanchez2016b}
Sanchez S.~F.,  et~al., 2016b, \rmxaa, 52, 171

\bibitem[\protect\citeauthoryear{Sanchez et~al.,}{Sanchez
  et~al.}{2018}]{Sanchez2018}
Sanchez S.~F.,  et~al., 2018, \rmxaa, 54, 217

\bibitem[\protect\citeauthoryear{Sanchez et~al.,}{Sanchez
  et~al.}{2019}]{Sanchez2019}
Sanchez S.~F.,  et~al., 2019, \mn@doi [\mnras] {10.1093/mnras/stz019}, 484,
  3042

\bibitem[\protect\citeauthoryear{Sanders et~al.,}{Sanders
  et~al.}{2020}]{Sanders2020}
Sanders R.~L.,  et~al., 2020, arXiv e-prints, p. arXiv:2009.07292

\bibitem[\protect\citeauthoryear{Scudder, Ellison  \& Mendel}{Scudder
  et~al.}{2012a}]{Scudder2012}
Scudder J.~M.,  Ellison S.~L.,   Mendel J.~T.,  2012a, \mn@doi [\mnras]
  {10.1111/j.1365-2966.2012.21080.x}, 423, 2690

\bibitem[\protect\citeauthoryear{Scudder, Ellison, Torrey, Patton  \&
  Mendel}{Scudder et~al.}{2012b}]{Scudder2012b}
Scudder J.~M.,  Ellison S.~L.,  Torrey P.,  Patton D.~R.,   Mendel J.~T.,
  2012b, \mn@doi [\mnras] {10.1111/j.1365-2966.2012.21749.x}, 426, 549

\bibitem[\protect\citeauthoryear{Smee et~al.,}{Smee et~al.}{2013}]{Smee2013}
Smee S.~A.,  et~al., 2013, \mn@doi [\aj] {10.1088/0004-6256/146/2/32}, 146, 32

\bibitem[\protect\citeauthoryear{Stasi{\'{n}}ska, Cid~Fernandes, Mateus,
  Sodr{\'e}~Jr.  \& Asari}{Stasi{\'{n}}ska et~al.}{2006}]{Stasinska2006}
Stasi{\'{n}}ska G.,  Cid~Fernandes R.,  Mateus A.,  Sodr{\'e}~Jr. L.,   Asari
  N.~V.,  2006, \mn@doi [\mnras] {10.1111/j.1365-2966.2006.10732.x}, 371, 972

\bibitem[\protect\citeauthoryear{Storchi-Bergmann, Calzetti  \&
  Kinney}{Storchi-Bergmann et~al.}{1994}]{StorchiBergmann1994}
Storchi-Bergmann T.,  Calzetti D.,   Kinney A.~L.,  1994, \mn@doi [\apj]
  {10.1086/174345}, 429, 572

\bibitem[\protect\citeauthoryear{Sutherland \& Dopita}{Sutherland \&
  Dopita}{1993}]{Sutherland1993}
Sutherland R.~S.,  Dopita M.~A.,  1993, \mn@doi [\apjs] {10.1086/191823}, 88,
  253

\bibitem[\protect\citeauthoryear{Sutherland, Dopita, Binette  \&
  Groves}{Sutherland et~al.}{2013}]{Sutherland2013}
Sutherland R.,  Dopita M.,  Binette L.,   Groves B.,  2013, ascl.soft, p.
  ascl:1306.008

\bibitem[\protect\citeauthoryear{Sutherland, Dopita, Binette  \&
  Groves}{Sutherland et~al.}{2018}]{Sutherland2018}
Sutherland R.,  Dopita M.,  Binette L.,   Groves B.,  2018, ascl.soft, p.
  ascl:1807.005

\bibitem[\protect\citeauthoryear{Teimoorinia, Jalilkhany, Scudder, Jensen  \&
  Ellison}{Teimoorinia et~al.}{2021}]{Teimoorinia2021}
Teimoorinia H.,  Jalilkhany M.,  Scudder J.~M.,  Jensen J.,   Ellison S.~L.,
  2021, \mn@doi [\mnras] {10.1093/mnras/stab466}, 503, 1082

\bibitem[\protect\citeauthoryear{Torrey, Cox, Kewley  \& Hernquist}{Torrey
  et~al.}{2012}]{Torrey2012}
Torrey P.,  Cox T.~J.,  Kewley L.,   Hernquist L.,  2012, \mn@doi [\apj]
  {10.1088/0004-637X/746/1/108}, 746, 108

\bibitem[\protect\citeauthoryear{Tremonti et~al.,}{Tremonti
  et~al.}{2004}]{Tremonti2004}
Tremonti C.~A.,  et~al., 2004, \mn@doi [\apj] {10.1086/423264}, 613, 898

\bibitem[\protect\citeauthoryear{Van Der~Walt, Colbert  \& Varoquaux}{Van
  Der~Walt et~al.}{2011}]{numpy}
Van Der~Walt S.,  Colbert S.~C.,   Varoquaux G.,  2011, \mn@doi [Computing in
  Science and Engineering] {10.1109/MCSE.2011.37}, 13, 22

\bibitem[\protect\citeauthoryear{Virtanen et~al.,}{Virtanen
  et~al.}{2020}]{scipy}
Virtanen P.,  et~al., 2020, \mn@doi [Nat Methods] {10.1038/s41592-019-0686-2},
  17, 261

\bibitem[\protect\citeauthoryear{Wake et~al.,}{Wake et~al.}{2017}]{Wake2017}
Wake D.~A.,  et~al., 2017, \mn@doi [\aj] {10.3847/1538-3881/aa7ecc}, 154, 86

\bibitem[\protect\citeauthoryear{Wright}{Wright}{2006}]{Wright2006}
Wright E.~L.,  2006, \mn@doi [\pasp] {10.1086/510102}, 118, 1711

\bibitem[\protect\citeauthoryear{Yan et~al.,}{Yan et~al.}{2016a}]{Yan2016a}
Yan R.,  et~al., 2016a, \mn@doi [\aj] {10.3847/0004-6256/151/1/8}, 151, 8

\bibitem[\protect\citeauthoryear{Yan et~al.,}{Yan et~al.}{2016b}]{Yan2016b}
Yan R.,  et~al., 2016b, \mn@doi [\aj] {10.3847/0004-6256/152/6/197}, 152, 197

\bibitem[\protect\citeauthoryear{York et~al.,}{York et~al.}{2000}]{York2000}
York D.~G.,  et~al., 2000, \mn@doi [\aj] {10.1086/301513}, 120, 1579

\bibitem[\protect\citeauthoryear{Zahid, Kewley  \& Bresolin}{Zahid
  et~al.}{2011}]{Zahid2011}
Zahid H.~J.,  Kewley L.~J.,   Bresolin F.,  2011, \mn@doi [\apj]
  {10.1088/0004-637X/730/2/137}, 730, 137

\bibitem[\protect\citeauthoryear{Zaritsky, Kennicutt  \& Huchra}{Zaritsky
  et~al.}{1994}]{Z94}
Zaritsky D.,  Kennicutt R. C.~J.,   Huchra J.~P.,  1994, \mn@doi [\apj]
  {10.1086/173544}, 420, 87

\bibitem[\protect\citeauthoryear{Zhang et~al.,}{Zhang et~al.}{2017}]{Zhang2016}
Zhang K.,  et~al., 2017, \mn@doi [\mnras] {10.1093/mnras/stw3308}, 466, 3217

\makeatother
\end{thebibliography}


\appendix

\section{Tables}
In this Appendix, we provide a sample few rows of the tables which list conversions between all calibrations, the number of spaxels used in the fitting procedure, the range of validity, and the polynomial fits used in this work as an example of the data structure. 
\begin{landscape}
\begin{table}
	\centering
	\caption{Summary of the median offset from a 5th order polynomial best fit in both positive and negative directions, for the contour that encloses 95.5 per cent of the data. For each metallicity calibration pairing, the number of spaxels which are present is also recorded, for the SMC dust correction curve. The full table, along with the same for all other pairings, and for the MW dust correction curve, is available as supplementary material.\label{tab:smc_values}}
\begin{tabular}{lccccccccccll}
\hline
Conversion $x \rightarrow y$ & $n$ spaxels & 2$\sigma$ scatter &  Range of validity & \multicolumn{6}{|c|}{Polynomial fit: a + b$x$ + c$x^2$ + d$x^3$ + e$x^4$ + f$x^5$} \\ 
&&& $\left(12+\text{log(O/H)} \right)$ & a & b & c & d & e & f \\ 
\hline 
 Z94 $\rightarrow$ PP04 O3N2 & 1047043 & [-0.1235, 0.1251] & [8.3647, 9.2955] & -8770.91616636 & + 5802.01495817 &-1482.76057393 & + 184.82094889 &-11.30013709 & + 0.27215961 \\ 
\hline 
 Z94 $\rightarrow$ M13 O3N2 & 1044153 & [-0.0826, 0.0827] & [8.3745, 9.3053] & -47272.33798346 & + 27298.59022885 &-6287.55001335 & + 722.2414351 &-41.3814117 & + 0.94623929 \\ 
\hline 
 Z94 $\rightarrow$ C17 O3N2 & 1047222 & [-0.092, 0.0873] & [8.3745, 9.3053] & -32271.51422548 & + 19287.44483188 &-4581.46297283 & + 541.17950728 &-31.80856677 & + 0.74459933 \\ 
\hline 
 M91 $\rightarrow$ Z94 & 1037616 & [-0.0264, 0.0206] & [8.4035, 9.0696] & -82822.51326969 & + 51125.46960079 &-12549.66580389 & + 1532.47593546 &-93.14893094 & + 2.25576348 \\ 
\hline 
 M91 $\rightarrow$ KK04 & 1055420 & [-0.0199, 0.0098] & [8.4035, 9.0712] &  + 49862.66651585 &-26473.36992983 & + 5590.08956813 &-586.27362519 & + 30.50799985 &-0.62930761 \\ 
 &  & \textit{lower branch:} & [7.8508, 8.2351] & -25.16773802 & + 7.44906684 &-0.40975953 &&&\\ 
\hline 
...&...& ...& ... &...&...& ...& ... &...&...\\
 \hline 
 \end{tabular}
\end{table}
\end{landscape}


\bsp	
\label{lastpage}
\end{document}